\documentclass[pageno]{jpaper}

\usepackage{amsmath,amssymb,amsfonts}
\usepackage[ruled,vlined,linesnumbered,resetcount,noend]{algorithm2e}
\usepackage{textcomp}
\usepackage{xcolor}
\usepackage[normalem]{ulem}
\usepackage{graphicx}
\usepackage{dirtytalk}

\usepackage[title]{appendix}
\usepackage{booktabs}
\usepackage{comment}
\usepackage{caption}
\usepackage{subcaption}
\usepackage{makecell}
\usepackage{svg}
\usepackage{tcolorbox}
\definecolor{aliceblue}{rgb}{0.94, 0.97, 1.0}
\usepackage{xcolor}
\tcbset{width=1\columnwidth,boxrule=1pt,colback=aliceblue,arc=1pt,left=2pt,right=2pt,boxsep=0.5pt}
\usepackage{enumitem}     
\usepackage{etoolbox}

\newcommand{\ignore}[1]{}

\usepackage{cite}
\usepackage[capitalise]{cleveref}

\usepackage[normalem]{ulem}

\newcommand{\framework}{{\'E}liv{\'a}gar}

\newcommand{\metricone}{representational capacity}
\newcommand{\Metricone}{Representational capacity}

\newcommand{\metrictwo}{Clifford noise resilience}

\newcommand{\MT}{CNR}
\newcommand{\MO}{RepCap}

\newcommand{\speedup}{271$\times$}
\newcommand{\perfgain}{5.3\%}
\newcommand{\perfgainhuman}{22.6\%}

\newcommand{\Sash}[1]{\textcolor{ForestGreen} {Sash: #1}}

\newcommand{\new}[1]{{#1}}
\newcommand{\imp}[1]{{#1}}
 
\definecolor{oldaliceblue}{rgb}{0.94, 0.97, 1.0}
\definecolor{aliceblue}{rgb}{0.14, 0.01, 0.8}
\definecolor{ForestGreen}{HTML}{007500}
\definecolor{orange}{HTML}{1A4D2E}
\definecolor{CoolPurple}{HTML}{7d70ba}
\definecolor{CoolGreen}{HTML}{0074E0}

\definecolor{titleblue}{HTML}{003076}
\definecolor{backgroundblue}{HTML}{eef6fc}

\newtcolorbox{hintbox}[2][]
{
  colframe = titleblue!100,
  colback  = backgroundblue!100,
  boxsep=2pt,
  width=\dimexpr\columnwidth\relax, 
  coltitle = titleblue!20!black,
  title    = #2,
  #1,
}

\begin{document}

\author{
  Sashwat Anagolum\thanks{ssa5517@psu.edu}\\
  Pennsylvania State University\\
  \and
  Narges Alavisamani\\
  Georgia Institute of Technology
  \and
  Poulami Das\\
  University of Texas, Austin
  \and
  Moinuddin Qureshi\\
  Georgia Institute of Technology
  \and
  Eric Kessler\\
  AWS Quantum Technologies
  \and
  Yunong Shi\\
  AWS Quantum Technologies
}

\title{\framework{}: Efficient Quantum Circuit Search for Classification}
\pagenumbering{arabic}

\date{}
\maketitle

\thispagestyle{empty}



\begin{abstract}
   Designing performant and noise-robust circuits for Quantum Machine Learning (QML) is challenging --- the design space scales exponentially with circuit size, and there are few well-supported guiding principles for QML circuit design. Although recent Quantum Circuit Search (QCS) methods attempt to search for performant QML circuits that are also robust to hardware noise, they directly adopt designs from classical Neural Architecture Search (NAS) that are misaligned with the unique constraints of quantum hardware, resulting in high search overheads and severe performance bottlenecks.

   We present \framework{}, a novel resource-efficient, noise-guided QCS framework. \framework{} innovates in all three major aspects of QCS --- search space, search algorithm and candidate evaluation strategy --- to address the design flaws in current classically-inspired QCS methods. \framework{} achieves hardware-efficiency and avoids an expensive circuit-mapping co-search via noise- and device topology-aware candidate generation. By introducing two cheap-to-compute predictors,  \textit {\metrictwo} and \textit {\metricone},  \framework{} decouples the evaluation of noise robustness and performance, enabling early rejection of low-fidelity circuits and reducing circuit evaluation costs. Due to its resource-efficiency, \framework{} can further search for data embeddings, significantly improving performance.

  Based on a comprehensive evaluation of \framework{} on 12 real quantum devices and 9 QML applications, \framework{} achieves \perfgain{} higher accuracy and a \speedup{} speedup compared to state-of-the-art QCS methods.
\end{abstract}
\vspace*{-0.1cm}

\section{Introduction}
\label{sec:intro}
  

Quantum Machine Learning (QML) is an important class of quantum algorithms for the \textit{Noisy-Intermediate Scale Quantum (NISQ)}~\cite{Preskill2018quantumcomputingin} era, due to its applicability to real-world problems such as classification \cite{wang2021quantumnas, farhi2018classification, lloyd2020quantum, nature_communication_2021_power_of_data_qml}, generative modeling \cite{huang2021qgans, neven2021entanglingqgans, ortiz2020iontrapqgans, zoufal2021qbms, zoufal2021thesis} and learning physical systems \cite{liu2021robustandrigorousspeedup, huang2022advantagefromexperiments}. QML uses \textit{variational} quantum circuits to perform machine learning tasks. The parameters in these variational circuits are iteratively updated (i.e., \textit{trained}) using a classical optimizer until the circuit learns the input problem to high accuracy. 

The key challenge in designing effective near-term QML algorithms is to find variational circuits with good \textit{circuit performance}, to accurately learn the input problem, and \textit{noise robustness}, to avoid performance degradation. However, due to the complexity of device-level noise sources and the nascent state of QML theory, there are few established guiding principles in designing QML circuits. Current manually crafted circuits are often under-performant, and face numerous practical issues such as vanishing gradients~\cite{wang2021noiseinduced} and low fidelity.

\begin{figure}[t]
        \centering
        \includegraphics[width=0.77\columnwidth]{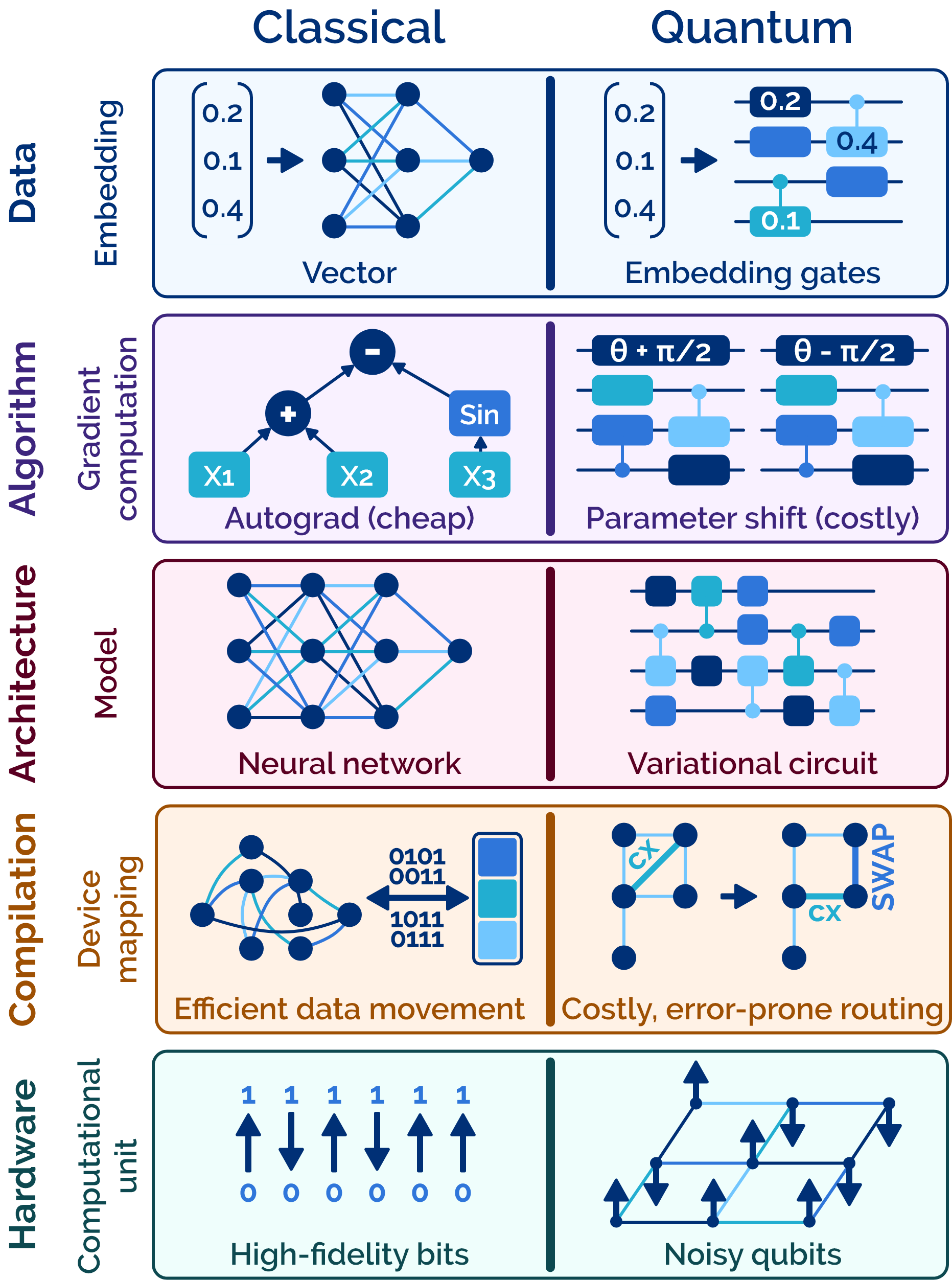}
        
        \caption{\new{Comparison of classical and quantum ML systems.}}
        \vspace*{-0.1cm}
        \label{fig:ml_comparison}
    \end{figure}





This challenge has motivated recent efforts in \textbf{Quantum Circuit Search (QCS)}~\cite{wang2021quantumnas, dacheng2020quantumsupernet, hong2020dqas} which aims to automatically find high-performance, noise-robust circuits. While promising, these methods are still preliminary, and naively adopt established methodologies from classical Neural Architecture Search (NAS)~\cite{li2019nassurvey, ning2020classicalnassurvey}, a subfield of Machine Learning (ML) that searches for high-performance neural networks. As a result, current QCS works overlook the fundamental differences between classical ML and QML, shown in~\cref{fig:ml_comparison}. 

 At the data processing level, classical ML uses multi-dimensional vectors to represent input data. In QML, data must be embedded into \textit{data embedding} gates, the choices and locations of which impact circuit performance. At the algorithm level, QML cannot compute gradients efficiently via automatic differentiation as intermediate states cannot be copied~\cite{nocloning} and collapse on measurement~\cite{NielsenChuang}. Instead, QML uses methods like parameter-shift rules~\cite{schuld2019analyticgradients} that scale poorly with problem size. At the compilation level, mapping classical neural networks with arbitrary topology onto memory is inexpensive due to efficient data movement. However, due to the limited connectivity of NISQ devices, large routing costs are incurred when QML circuits do not match device topology. 
 
\begin{figure}[t]
        \centering
        \includegraphics[width=0.73\columnwidth]{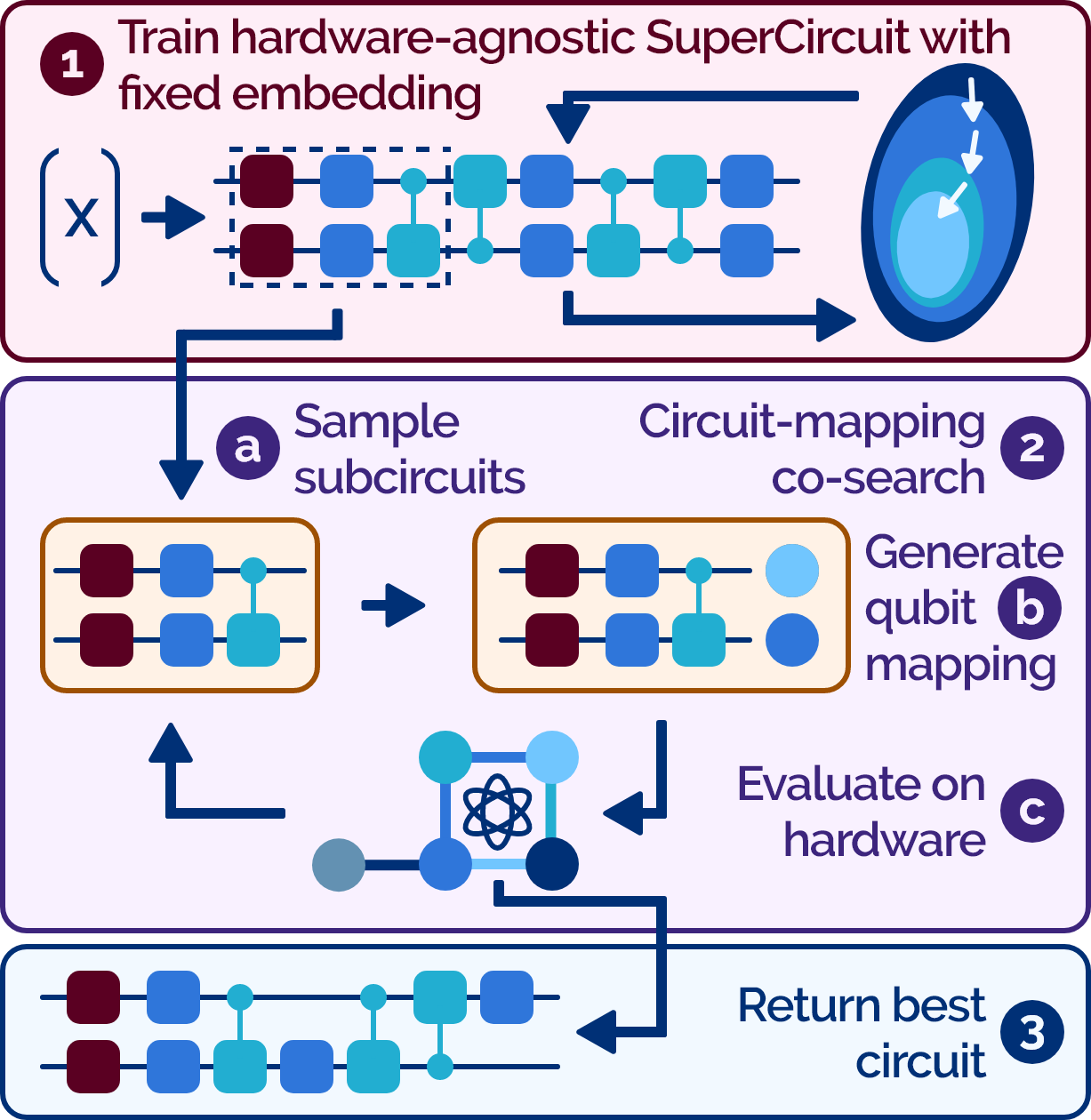}
        
        \caption{QuantumNAS: A state-of-the-art QCS framework.}
        \vspace*{-0.3cm}
        \label{fig:quantumnas}
\end{figure}

Existing QCS methods naively adopt classical ML designs, precluding compatibility with the unique constraints of QML. We explain these mismatches using QuantumNAS~\cite{wang2021quantumnas}, a state-of-the-art QCS method. As shown in ~\cref{fig:quantumnas}, QuantumNAS first 
trains a large device-agnostic \emph{SuperCircuit} with fixed data embeddings. Next, it performs an evolutionary co-search to identify a performant and noise-robust circuit-qubit mapping pair, using the trained SuperCircuit parameters to estimate the performance of candidate circuits on the target device. Finally, it returns the best circuit-qubit mapping pair found. Other frameworks such as QuantumSupernet~\cite{dacheng2020quantumsupernet} follow similar designs, and thus also experience the following issues:

\begin{enumerate}[leftmargin=0cm,itemindent=.5cm,labelwidth=\itemindent,labelsep=0cm,align=left, itemsep=0.2cm, listparindent=0.4cm]
    \item They use device-agnostic search spaces, necessitating an additional search for logical-to-physical qubit mappings. The found mappings are also hardware-inefficient, since selected circuits usually do not match the device topology and SWAP gates must be inserted before circuit execution.

    \item They do not assess the impact of data embeddings on circuit performance as they use fixed, dataset-agnostic data embeddings. This leads to circuits with poor performance.

    \item They adopt strategies from classical ML relying heavily on gradient computation, which is expensive on quantum computers and scales poorly with circuit size. 
    
    \item They perform expensive performance evaluation for all circuits. This is inefficient, since many circuits have extremely low fidelity, and will perform poorly since their outputs will be severely corrupted by to device noise. Existing methods do not have any mechanisms to identify and eliminate the performance evaluation for such circuits. 
\end{enumerate}

Consequently, existing QCS frameworks incur significant search overheads even for small circuits and problem sizes, and often perform poorly on QML tasks.

To address the problems above, we propose \framework{}, a resource-efficient, high-performance QCS framework tailored to QML systems. \textbf{First}, \framework{} uses a device topology- and noise-aware circuit generation process to generate noise robust circuits that are hardware efficient, eliminating the costly circuit-mapping co-search (issue 1). \textbf{Second}, to mitigate performance bottlenecks caused by fixed, data-agnostic data embeddings (issue 2), \framework{} generates circuits with different data embeddings and variational gates, allowing \framework{} to search for the optimal data embedding for a QML task.
\textbf{Third}, to reduce the overheads of training-based circuit evaluations (issue 3), \framework{} introduces a novel, cheap-to-compute performance predictor, \textit{\metricone{}}, for circuit performance evaluation. By using \metricone{}, \framework{} eliminates the expensive training step in the QCS workflow while still effectively predicting circuit performance. \textbf{Last}, \framework{} makes use of the insight that evaluating noise robustness is much simpler than evaluating performance. By introducing \textit{\metrictwo{}}, a predictor of circuit noise robustness, it decouples these two evaluations to enable early rejection of low-fidelity circuits (issue 4). 


Our evaluations using 9 near-term QML benchmarks show that \framework{} finds circuits with \perfgain{} higher accuracy while being \speedup{} faster than prior state-of-the-art QCS methods. Moreover, we show that the speedup of \framework{} increases with problem size. In summary, our contributions are:


\begin{enumerate}[leftmargin=0cm,itemindent=.5cm,labelwidth=\itemindent,labelsep=0cm,align=left, itemsep=0.2cm, listparindent=0.4cm]
    \item  We propose a novel QCS workflow, \framework{}, that is tailored to QML systems and addresses the design flaws in current classically-inspired QCS methods. 
    
    \item We show that noise-guided, topology-aware and data embedding-aware circuit search, on average, leads to a \perfgain{} improvement in circuit performance compared to current methods.

    \item We show that by using \metricone{} to predict circuit performance, the number of circuit executions in \framework{} can be improved by 22$\times$- 523$\times$ relative to training-based evaluation strategies.

    \item We show that by performing early rejection of low-fidelity circuits using \metrictwo{}, the resource efficiency of \framework{} can be further improved by 2$\times$-20$\times$ depending on noise level.

    
\end{enumerate}

\new{We open source \framework{} to facilitate future QCS research at \href{https://github.com/SashwatAnagolum/Elivagar}{https://github.com/SashwatAnagolum/Elivagar}.}

\section{Background and Motivation}
\label{sec:background}
\subsection{Executing quantum programs on NISQ computers}
\label{sec:nisqintro}

\begin{figure}[t]
     \centering
     \begin{subfigure}{1\columnwidth}
         \centering
         \includegraphics[width=0.7\columnwidth]{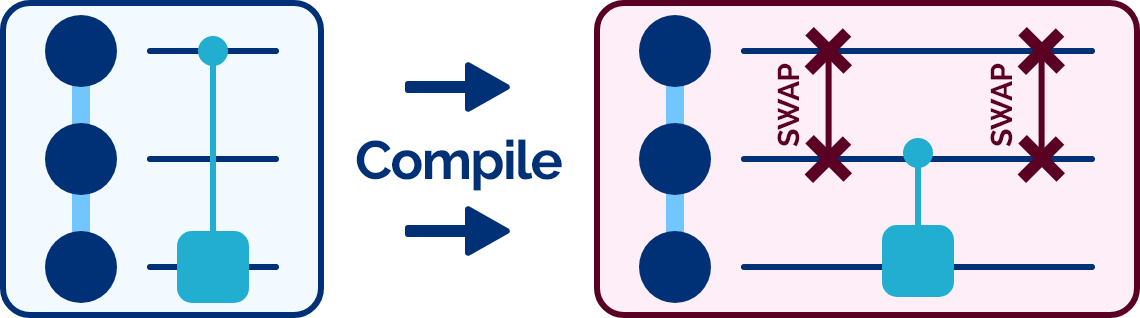}
         \caption{}
         \label{fig:swap}
     \end{subfigure}
     \hfill
     \begin{subfigure}{1\columnwidth}
         \centering
         \includegraphics[width=0.8\columnwidth]{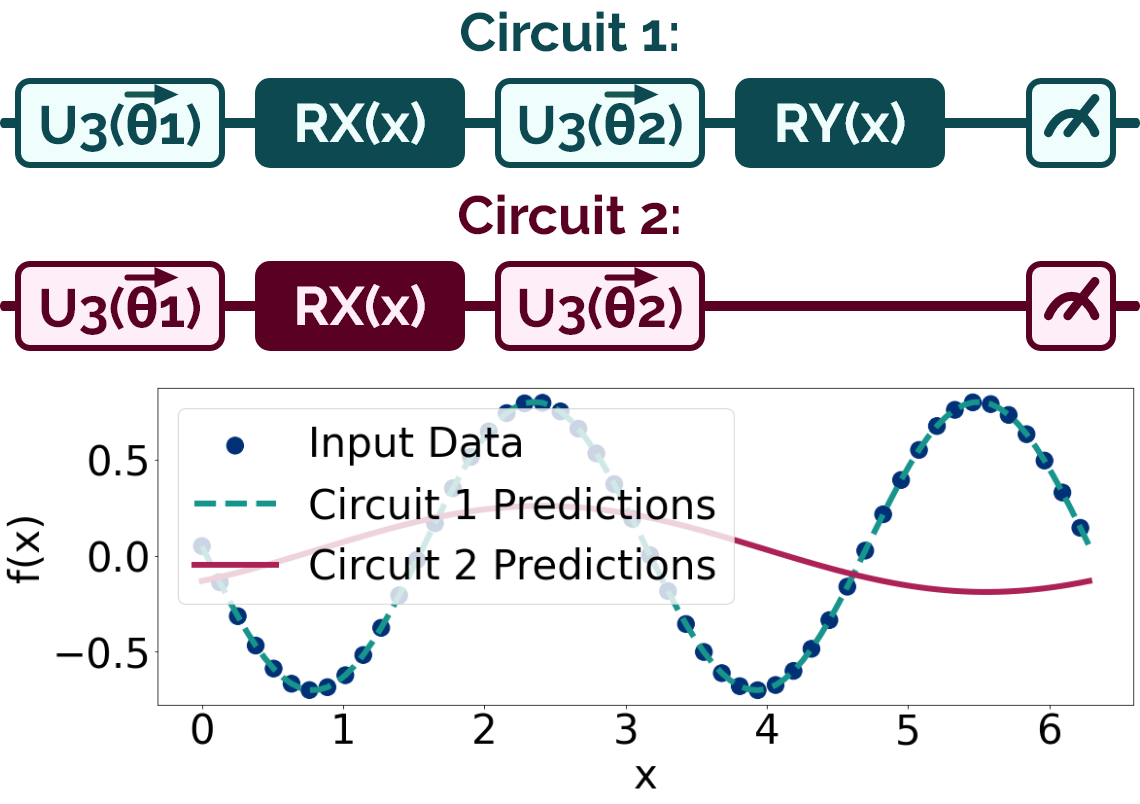}
         \caption{}
         \label{fig:embedding}
     \end{subfigure}
     \hfill
     \begin{subfigure}{1\columnwidth}
         \centering
         \includegraphics[width=0.9\columnwidth]{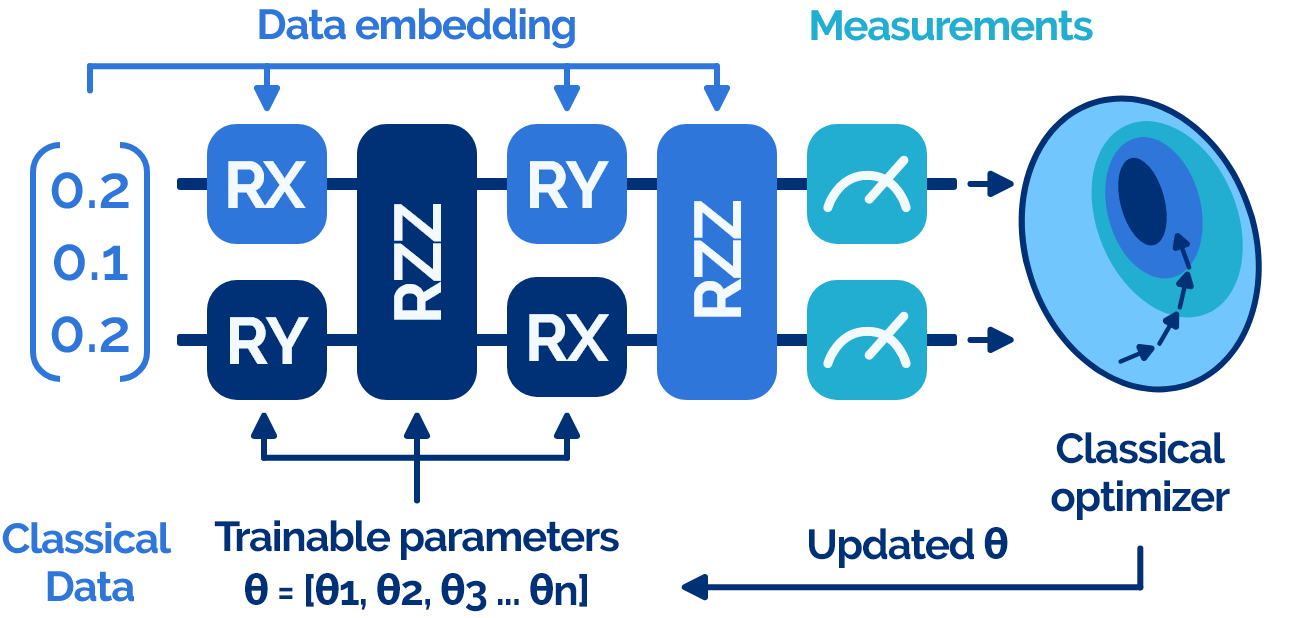}
         \caption{}
         \label{fig:qml_training}
     \end{subfigure}
     \vspace*{-0.4cm}
     \caption{(a) Circuit and device topology mismatch results in a large routing cost due to noisy SWAP gates. \new{(b) Both circuits use the same trainable gates, but different data embeddings. Circuit 1 learns the target function $f(x)$ but Circuit 2 fails, highlighting the importance of suitable data embeddings for QML tasks. (c) QML circuits consist of data embedding gates, trainable gates, and measurement operations. QML training involves running circuits on a quantum device and tuning parameters using a classical optimizer.}}
     \vspace*{-0.4cm}
\end{figure}

NISQ-era devices with limited qubit connectivity and imperfect operations make program executions error-prone. Operations on non-adjacent qubits are enabled using SWAPs, as shown in~\cref{fig:swap}, which are usually implemented using three 2-qubit gates, making them noisy and expensive. Prior works~\cite{murali2020crosstalk, das2021adapt, ravi2022cliffordcanarycircuits, tannu2022hammer, li2018sabre} show that circuit fidelity can be improved through circuit compilation techniques such as \emph{qubit mapping and routing}, which minimize SWAPs, making it desirable to eliminate SWAPs to make circuits hardware-efficient and boost circuit fidelity.

\renewcommand\theadalign{bc}
\renewcommand\theadfont{\bfseries}
\renewcommand\theadgape{\Gape[6pt]}
\renewcommand\cellgape{\Gape[6pt]}
\begin{table*}[t]
    \centering
    \setlength{\tabcolsep}{0.26cm}
    \caption{\new{Key differences between \framework{} and existing QCS methods. ES, RL, and MCTS refer to evolutionary search (ES), reinforcement learning (RL), and Monte Carlo tree search (MCTS), respectively.}}
    \begin{tabular}{l c c c c c }
        \toprule
       Framework & \makecell{Search\\Space} & \makecell{Search\\Algorithm} & \makecell{Candidate\\Evaluation Strategy} & \makecell{Runtime\\Bottleneck}\\
        \midrule
        MCTS-QAOA~\cite{yao2022mctsqaoa} & \makecell{Pauli strings} & \makecell{RL + MCTS} & \makecell{Train circuit} & \makecell{Sample-inefficient RL}\\

        QCEAT~\cite{huang2022qceat} & \makecell{Random circuits} & \makecell{ES} & \makecell{Train circuit} & \makecell{Gradient computation}\\
        
        QuantumSupernet~\cite{dacheng2020quantumsupernet} & \makecell{SuperCircuit} & \makecell{Random search} & \makecell{SuperCircuit loss} & \makecell{Gradient computation}\\
        
        QuantumNAS~\cite{wang2021quantumnas} & \makecell{SuperCircuit} & \makecell{ES} & \makecell{SuperCircuit loss} & \makecell{Gradient computation}\\
        
        \midrule
        \framework{} & \makecell{Device- and noise-\\aware circuits} & \makecell{Noise-guided random\\search (Sec. \ref{sec:circuit_generation})} & \makecell{\MT{} (Sec. \ref{sec:clifford_noise})\\and \MO{} (Sec. \ref{sec:metrics_explanations})} & \makecell{--}\\
        \bottomrule
        \end{tabular}
    \label{tab:qcs_comparison}
\end{table*}
 
\subsection{Variational circuits for QML}
\label{sec:qml_intro}

Variational circuits in QML consist of data embedding gates, trainable variational gates, and measurement operations. The data embedding gates are used to transform classical data into quantum states by using these classical data as rotation angles for the data embedding gates. Once the data has been embedded, the variational gates in the circuit are used to manipulate the representations of the data, and measurements are applied by using measurement operators to extract classical data for post-processing, such as for computing parameter updates during training, or for making predictions during inference.


After training, a QML circuit can be used for inference. To do so, we embed into the circuit the data to make predictions based on, and use the learned parameter values for the variational gates. The circuit is executed on a quantum device, and measurements are performed to extract classical data which can be post-processed to obtain predictions. The post-processing can potentially involve statistical analysis, decoding, or other task-specific operations transforming measurement outputs into meaningful results.

Different circuits can be constructed by varying the number, type, and placement of gates used. Due to the large search space, designing performant QML circuits is challenging. As a result, practitioners often rely on a set of commonly used variational \emph{templates}, such as the hardware efficient ansatz \cite{mcclean2018barrenplateaus}. However, numerous works~\cite{cerezo2021costfndepbarrenplateaus, nature_communication_2021_power_of_data_qml, mcclean2018barrenplateaus, canatar2022bandwidthinquantumkernels} show that these templates tend to perform poorly on QML tasks as circuit size increases, motivating the search for better circuit structures.

\subsubsection{Data embedding gates}
\label{sec:data_embedding}
The choice of data embedding in a QML circuit significantly impacts its performance, as highlighted in prior studies \cite{schuld2022fourierinterpolation, schuld2021encodingexpressivity, schuld2021qmlmodelsarekernelmethods, sweke2021encodingdependentbounds}. To illustrate, consider the example in \cref{fig:embedding}, where two circuits share the same variational gates, but differ in data embeddings (RX$(x)$ and RY$(x)$, versus RX$(x)$ alone), resulting in markedly different performance outcomes. \new{However, choosing a suitable data embedding for a given QML task is challenging due to the vast search space, leading many QML circuits to use a fixed embedding such as an angle \cite{qml} or Instantaneous Quantum Polynomial-time (IQP) embedding \cite{havlek2019suplearninginquantumenhancedfeaturespaces}, regardless of the nature of the QML task.} This often results in suboptimal performance due to the mismatch between the embedding used and the QML task.

\subsubsection{Training variational circuits}
\label{sec:training_qml}

Variational circuits are trained using gradient-based optimization in a classical-quantum feedback loop, as shown in \cref{fig:qml_training}. However, training variational circuits is costlier than training classical ML models. In classical ML, gradients are efficiently computed via Automatic Differentiation (AD)\cite{ad_survey}, which takes constant time. However, this approach is not feasible on quantum computers since intermediate quantum states cannot be manipulated or copied due to the no-cloning theorem \cite{NielsenChuang, nocloning}. Thus, alternative methods such as parameter-shift rules \cite{schuld2019analyticgradients, wierichs2022parametershift} are used. Parameter-shift methods compute the gradient of a single parameter $\theta$ by running two circuits with shifted parameters, $\theta+s$ and $\theta-s$ (where $s$ is a constant), and then computing the difference. Unfortunately, this method requires circuit executions scaling linearly with the number of circuit parameters, in contrast to the constant scaling of AD methods; thus, QML gradient computation scales poorly with the number of parameters.


        

\subsection{Limitations of existing QCS works}
\new{QCS works that use reinforcement learning (RL), such as MCTS-QAOA \cite{yao2022mctsqaoa}, converge slowly due to the large action and state spaces involved in quantum circuit design, and incur large circuit evaluation costs during training of the RL model.} SuperCircuit-based works such as QuantumSupernet~\cite{dacheng2020quantumsupernet} and QuantumNAS \cite{wang2021quantumnas} also result in impractical runtime overheads due to the high cost of SuperCircuit training (\cref{sec:training_qml}). These methods additionally require a circuit-mapping co-search which is often unable to make circuits hardware-efficient, leading to SWAP insertions and reduced circuit noise robustness (\cref{sec:nisqintro}). Moreover, SuperCircuit-based works cannot search for optimal data embeddings for a QML task, leading to lowered performance due to data embedding-QML task mismatch (\cref{sec:data_embedding}). Together, these issues hinder the ability of existing works to find performant circuits while maintaining low search costs.

\subsection{The necessity of a new QCS pipeline}
It is infeasible to improve significantly on existing QCS works since many of the prerequisites for performant, low-cost QCS are fundamentally incompatible with SuperCircuit-based methods. For example, SuperCircuit training is necessary for accurate circuit performance prediction, preventing the elimination of training and gradient computation. However, gradient computation on quantum hardware is costly, leading to high search overheads. Additionally, SuperCircuit performance prediction is only accurate when using a fixed data embedding for all candidate circuits, precluding the possibility of searching for optimal data embeddings to boost performance.

Therefore, to enable efficient QCS, it is necessary to develop a novel QCS pipeline that accounts for the constraints of QML systems and NISQ-era quantum hardware. To this end, we propose \framework{}, a QCS framework which \new{eschews} using a SuperCircuit in favor of training-free predictors of circuit performance, avoiding the training-induced runtime bottlenecks of prior work. Moreover, since \framework{} does not use a SuperCircuit, it can search for optimal data embeddings and perform efficient noise-guided candidate sampling, allowing it to generate hardware-efficient circuits at neglible cost and outperform prior QCS works consistently. \framework{}'s key differences with prior QCS works are presented in \cref{tab:qcs_comparison}.

\begin{figure*}[t]
    \centering
    \includegraphics[width=0.95\textwidth]{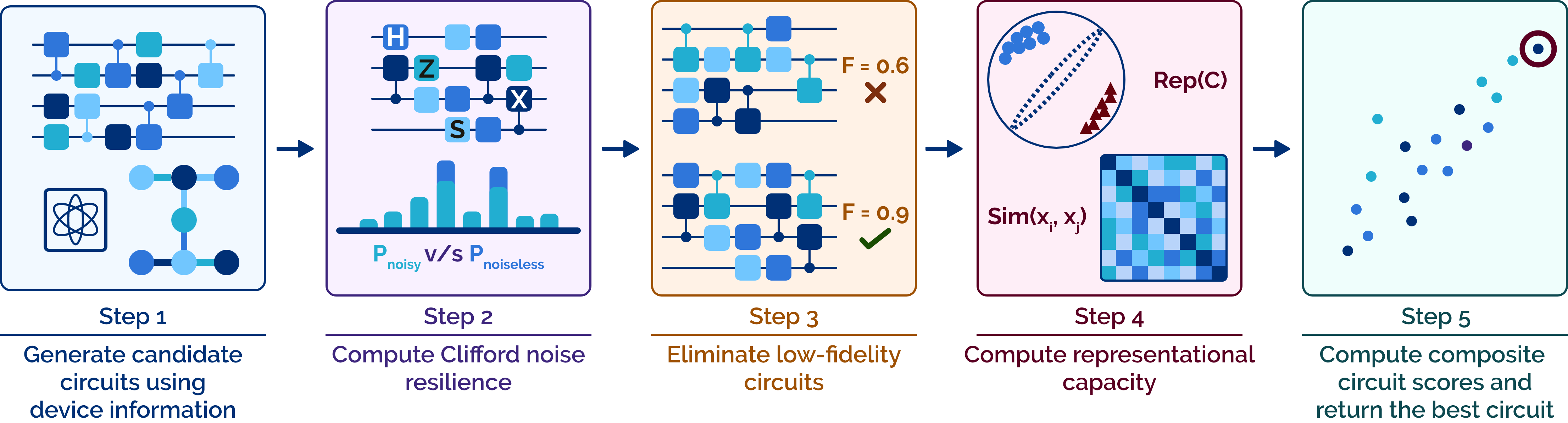}
        
    \caption{\new{An overview of \framework{}.}}
    \vspace*{-0.3cm}
    \label{fig:framework_overview}
\end{figure*}

\section{Overview of \framework{}}

As illustrated in~\cref{fig:framework_overview}, \framework{} consists of five steps:

\begin{enumerate}[leftmargin=0cm,itemindent=.5cm,labelwidth=\itemindent,labelsep=0cm,align=left, itemsep=0.2cm, listparindent=0.4cm]
    \item \textbf{Topology- and noise-aware candidate circuit generation:} First, \framework{} generates candidate circuits and data embeddings in a device- and noise-aware manner.
    \item \textbf{Evaluation of noise robustness:}  \framework{} then computes the \metrictwo{} score for circuits to estimate circuit noise robustness.
    \item \textbf{Early rejection of low-fidelity circuits:} \framework{} ranks candidate circuits based on their \metrictwo{} and eliminates low-fidelity circuits.
    \item \textbf{Evaluation of circuit performance:} \framework{} predicts the performance of the remaining circuits on the target QML task using \metricone{}.
    \item \textbf{Final circuit selection:} Finally, \framework{} computes composite scores combining \metricone{} and \metrictwo{} for the remaining candidate circuits, and returns the best circuit.
\end{enumerate}

We break \framework{} down into three parts --- candidate circuit generation (step 1, ~\cref{sec:circuit_generation}), noise-guided candidate rejection (step 2-3, ~\cref{sec:clifford_noise}), and circuit performance estimation and circuit selection (step 4-5, ~\cref{sec:metrics_explanations}).

\section{Candidate circuit generation}
\label{sec:circuit_generation}

To overcome the issues of SuperCircuit-based circuit generation discussed in~\cref{sec:intro}, \framework{} directly generates candidate circuits and data embeddings based on target device topology via a noise-guided sampling policy shown in~\cref{alg:candidategen}. This approach has the following advantages over SuperCircuit-based methods: 1) generating circuits based on device topology ensures hardware-efficiency and improves circuit fidelity; 2) circuits are generated along with optimal qubit mappings, eliminating the need for an expensive circuit-mapping co-search; 3) noise-guided sampling further boosts the average fidelity of candidate circuits; 4) generating varied data embeddings for candidate circuits improves selected circuit performance. 

As a result, circuits generated by \framework{} have 18.9\% higher fidelity than device-unaware circuits optimized using SABRE~\cite{li2018sabre} (\cref{sec:device_aware_gen}). Furthermore, \framework{} obtains 6\% higher accuracy on average when searching for data embeddings than when using a fixed data embedding (\cref{sec:accuracy_breakdown}).

\begin{algorithm}[t]
\SetAlgoLined
 
    \KwIn{Device topology graph ${G}(\{V_i\}, \{E_j\})$; Qubit readout errors $\{R_i\}$; Qubit coherence times $\{\mathrm{T1}_i \}, \{\mathrm{T2}_i\}$; \new{2-qubit gate fidelity $\{Q_e\}$}; Output circuit configuration $O_{\text{conf}}$.}
    \KwOut{Candidate circuit $C$.}

    Sample a set of connected subgraphs $\{S_i\}$ with $V^{(S_i)} = O_{\text{conf}}.n_q \: \forall_i$ from $G$;
    
    Select a subgraph $S(V^{(S)}, E^{(S)})$ from distribution 
    $\textrm{P}_{\{S_i\}}(S \: | \: R_{V^{(S)}}, \mathrm{T1}_{V^{(S)}}, \mathrm{T2}_{V^{(S)}}, Q_{E^{(S)}})$\;

    Initialize $C$ as an empty circuit\;

    Randomly sample a list $L_{\text{op}}$ of 1-qubit and 2-qubit gates based 
    on $O_{\text{conf}}.n_{\text{params}}$ and $O_{\text{conf}}.n_{\text{embeds}}$\;
 
    \For{$op$ in $L_{\text{op}}$}{
        \If{$op$ is a 1-qubit gate}{
            Select a qubit $q \in V^{(S)}$ from distribution $\textrm{P}_{1\text{q}}(q \: | \: C, \mathrm{T1}_{q}, \mathrm{T2}_{q})$\;
           
            Append $op(q)$ to $C$\;
         }
         
        \If{$op$ is a 2-qubit gate}{
            Select an edge $e = (q_0, q_1) \in E^{(S)}$ from $\textrm{P}_{2\text{q}}(e \: | \: C, \mathrm{T1}_{q_0}, \mathrm{T1}_{q_1}, \mathrm{T2}_{q_0}, \mathrm{T2}_{q_1}, Q_e)$\;
              
            Append $op(q_0, q_1)$ to $C$\;
        }
    }

    Select a set $q_{\text{meas}}$ of $O_{\text{conf}}.n_{\text{meas}}$ qubits in $V^{(S)}$ from distribution $\textrm{P}_{\text{meas}}(q \: | \: R_{E^{(S)}})$\;
    
    Append MEASURE($q_{\text{meas}}$) to $C$\;
    
    Randomly designate $O_{\text{conf}}.n_{\text{embeds}}$ parametric gates in $C$ for use as data embedding gates\;
    
    \Return $C$;
    \caption{Generate device-aware circuits}
    \label{alg:candidategen}
    \vspace{-4pt}
\end{algorithm}

\subsection{Noise-guided candidate circuit generation}
As shown in~\cref{alg:candidategen}, \framework{} first chooses a connected subgraph from the device topology, and then samples a list of gates to form the circuit. Gate placement choices are guided by subgraph connectivity, noise information such as qubit coherence times, 1- and 2-qubit gate fidelities, readout fidelities, and the current circuit structure. Following prior works in classical NAS~\cite{liu2018darts, yang2019gdas}, \framework{} samples qubits and gate placements from probability distributions instead of directly choosing the best options to encourage candidate diversity.

Choosing a subgraph $S(V^{(S)}, E^{(S)})$ for every circuit allows \framework{} to trivially obtain a qubit mapping for every generated circuit by using $V^{(S)}$. This way, \framework{} avoids the expensive circuit-mapping co-search required by other QCS frameworks. In contrast, the evolutionary co-search used by QuantumNAS~\cite{wang2021quantumnas} cannot guarantee hardware-efficiency, and is very expensive: eliminating it results in \framework{} becoming 1.4$\times$-33.4$\times$ faster (\cref{sec:source_of_speedup}). Thus, rather than use an alternative co-search mechanism, \framework{} adopts a novel approach that completely eliminates the need for a co-search, resulting in higher circuit fidelity and search efficiency.

\vspace{-0.1cm}
\begin{hintbox}{\color{white}\textbf{Insight 1: Generating device-aware circuits}}
Generating device- and noise-aware circuits eliminates the need for an expensive circuit-mapping co-search, improves circuit fidelity, and ensures hardware-efficiency.  
\end{hintbox}

\begin{figure*}[t]
    \centering
     \begin{subfigure}{0.14\textwidth}
        \includegraphics[width=1\columnwidth]{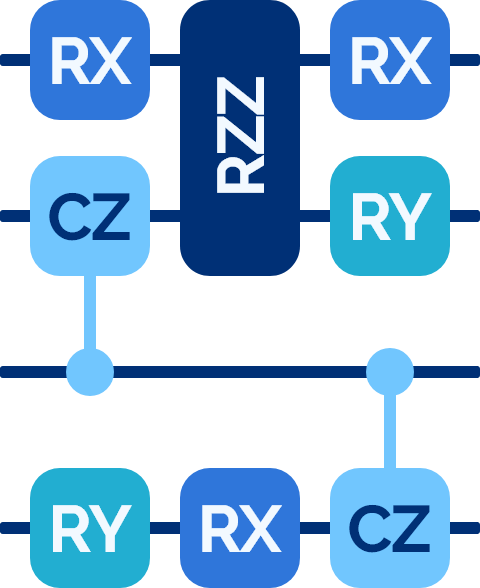}
       \caption{}
        \label{fig:original_circuit_sec_5}
    \end{subfigure}
        \hfill
     \begin{subfigure}{0.14\textwidth}
\includegraphics[width=1\columnwidth]{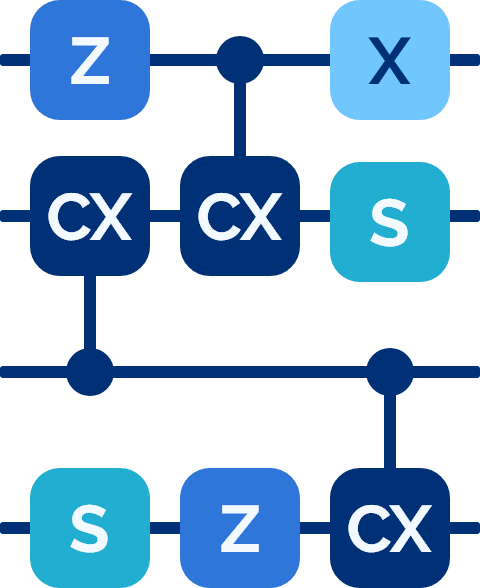}
       \caption{}
        \label{fig:clifford_circuit_sec_5}
    \end{subfigure}
        \hfill
         \begin{subfigure}{0.31\textwidth}
        \centering
        \includegraphics[width=1\textwidth]{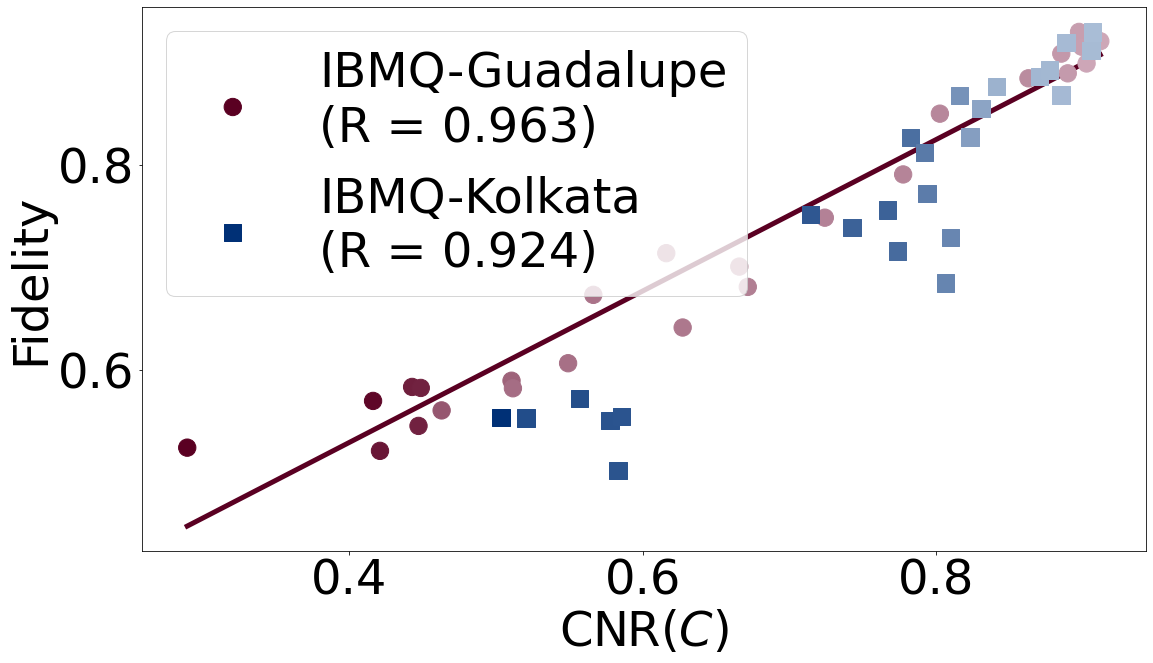}
        \caption{}
        \label{fig:clifford_correlation_realdevice}   
    \end{subfigure}
     \hfill
         \begin{subfigure}{0.31\textwidth}
        \centering
        \includegraphics[width=1\textwidth]{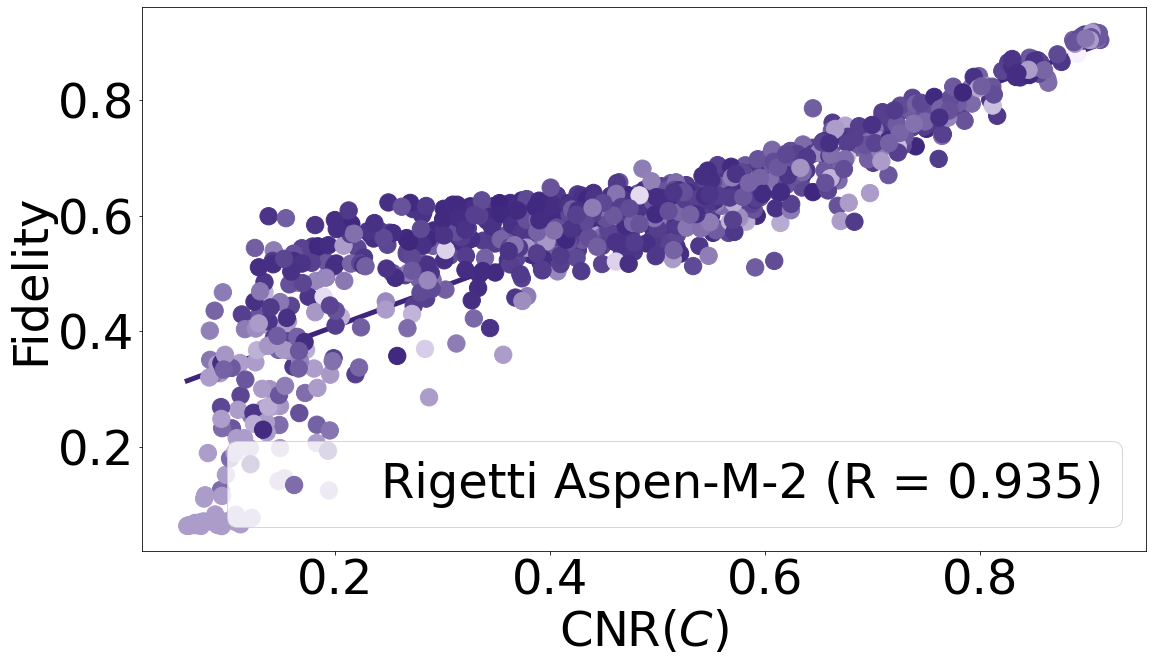}
        \caption{}
        \label{fig:clifford_correlation_simulator}   
    \end{subfigure}
    \vspace*{-0.3cm}
    \caption{(a) A circuit and (b) a Clifford replica of the circuit. \MT{} is strongly correlated with circuit fidelity, as demonstrated using circuits run on (c) IBMQ-Kolkata and IBMQ-Guadeloupe, and a (d) noise model of Rigetti Aspen-M-2.}
    \label{fig:correlations}
    \vspace*{-0.1cm}
\end{figure*}

\subsection{Generating data embeddings}
\label{sec:data_embeddings_strategy}

\framework{} employs a simple strategy to generate data embeddings: a few gates in each circuit are used as data embedding
gates, with each gate randomly assigned one dimension of the input data to embed (line 14 in~\cref{alg:candidategen}). Despite the random generation of data embeddings, \framework{} leverages \metricone{} (\cref{sec:metrics_explanations}) to predict circuit performance with different data embeddings and select circuits that contain the most suitable data embeddings for the QML task of interest. Consequently, \framework{} optimizes both the data embedding and the variational gates in the selected circuit, resulting in improved performance. By co-searching for data embeddings and variational gates, \framework{} unlocks 6\% higher accuracy than when using a fixed data embedding (\cref{sec:accuracy_breakdown}).

\begin{hintbox}{\color{white}\textbf{Insight 2: Searching for optimal data embeddings}}
Using optimized data embeddings significantly boosts performance over QCS methods with fixed embeddings.
\end{hintbox}

\section{Noise-guided candidate rejection}
\label{sec:clifford_noise}

A key insight of \framework{} is that predicting circuit noise robustness is much simpler than predicting circuit performance. Leveraging this, \framework{} reduces the number of performance estimations required by identifying and discarding low-fidelity circuits. To do so, \framework{} defines a predictor, \textit{\metrictwo{}} (\MT{}), that accurately predicts circuit fidelity. 

While simulating a quantum circuit is exponentially costly in general, Clifford circuits are a class of efficiently simulable quantum circuits~\cite{PhysRevLett_cliffordSimulation}. Thus, the fidelity of Clifford circuits can be calculated efficiently by comparing their outputs with noiseless classical simulation results. This enables efficient fidelity estimation for a circuit by using the fidelity of its \emph{Clifford replicas} as a proxy~\cite{Knill_2008_randomizedBenchmarking}, which is expensive to compute directly for large circuits due to the high simulation cost. While prior works use Clifford replicas to create circuit compilation passes~\cite{das2021adapt, das2022imitationgame, QAFCA} and characterize device noise~\cite{ravi2022cliffordcanarycircuits}, \framework{} uses Clifford circuits to identify noise-robust circuits for QML. First, we elaborate on Clifford replicas and \MT{}, and then show the strong correlation between \MT{} and circuit fidelity.

\subsection{Creating Clifford replicas} 
Clifford replicas of a circuit are created by substituting all of the non-Clifford gates in the circuit with Clifford gates. Previous works such as \cite{das2021adapt, das2022imitationgame} generate a Clifford replica for a circuit by substituting non-Clifford gates with the closest Clifford gate as measured by the diamond norm. In contrast, \framework{} uses multiple Clifford replicas with randomly chosen Clifford gates. The primary reason for this is that prior works deal with circuits at the compilation stage, where the parameter values to be used with each gate are known. In contrast, since \framework{} generates Clifford replicas for circuits before training, it is impossible to know what parameter values will be used with the circuit. Moreover, since gate angles will be changed due to training or different input data, using multiple Clifford replicas, each with randomly chosen Clifford gates, provides a reliable measure of circuit noise robustness over the course of training and prediction.

Figure \ref{fig:original_circuit_sec_5} and \ref{fig:clifford_circuit_sec_5} show a circuit and one of its Clifford replicas. Since Clifford replicas use the same structure as the original circuit, circuit properties such as depth, are preserved. \framework{} creates multiple Clifford replicas and averages the fidelities of all the constructed replicas. We find that using as few as 16 Clifford replicas can accurately characterize circuit noise robustness.

\subsection{Computing CNR} To compute the \MT{} for a circuit, \framework{} first collects the noisy outputs $P_{\textrm{noisy}}$ obtained by executing the generated Clifford replicas on the target quantum device. Then, it obtain the ideal Clifford replica outputs $P_{\textrm{noiseless}}$ via simulation. \framework{} then calculates the Total Variation Distance (TVD) between the noisy and noiseless outputs and computes the fidelity as $\textrm{Fid} = 1- \textrm{TVD}$: 

\begin{equation}
\label{eq:tvd}
\textrm{TVD} = \frac{1}{2} \sum_{i \in [2^{n_q}]} | P_{\textrm{noiseless}}(|i\rangle) - P_{\textrm{noisy}}(|i\rangle) |.
\end{equation}
$\textrm{CNR}(C)$, the \metrictwo{} for circuit $C$, is defined as the average fidelity of the $M$ Clifford replicas $C_{R}(i)$:

\begin{equation}
    \textrm{CNR}(C) = \frac{1}{M}\sum ^{M}_{i=1} \textrm{Fid}(P^{C_{R}(i)}_{\textrm{noiseless}},\: P^{C_{R}(i)}_{\textrm{noisy}}). 
    \label{eq:cnr}
\end{equation}

\subsection{Candidate circuit rejection}
After computing the \MT{} of candidate circuits, \framework{} ranks circuits by CNR and by default rejects circuits with \MT{} values less than a threshold of 0.7, or outside the top 50\% of all candidates. However, these values can be set by users of \framework{} as hyperparameters. All rejected circuits require no performance evaluation, leading to reduced resource requirements. Depending on device noise levels and threshold values, this rejection step can speed up \framework{} significantly. For example, when searching for a 250-parameter circuit on IBMQ-Manila with a \MT{} threshold of 0.9, \framework{} can reject 95\% of circuits, achieving an almost 20x reduction in circuit executions.

\begin{figure*}[t]
    \centering
     \begin{subfigure}{0.5\textwidth}
        \includegraphics[width=\columnwidth]
        {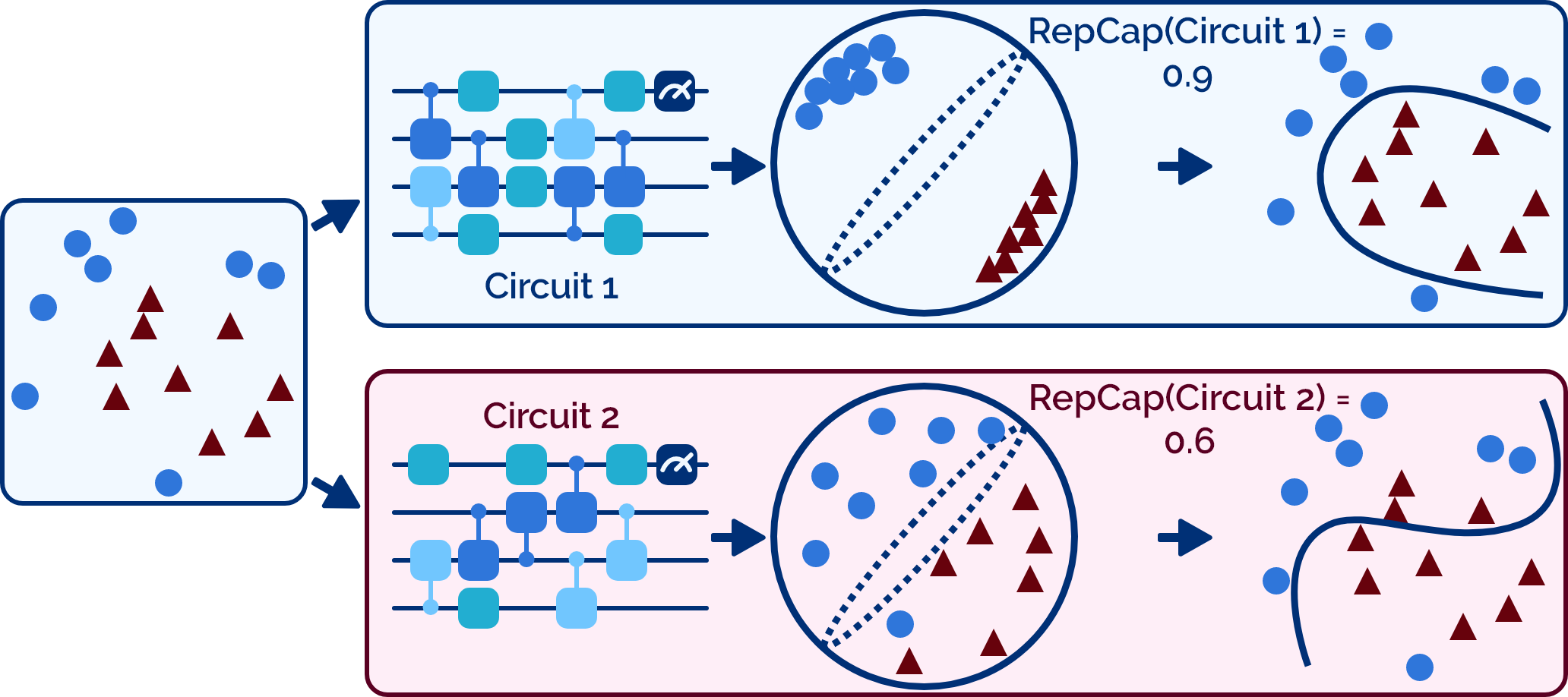}
       \caption{}
        \label{fig:rep_cap_diag}
    \end{subfigure}
        \hfill
    \begin{subfigure}{0.48\textwidth}
        \centering
        \includegraphics[width=0.49\columnwidth]{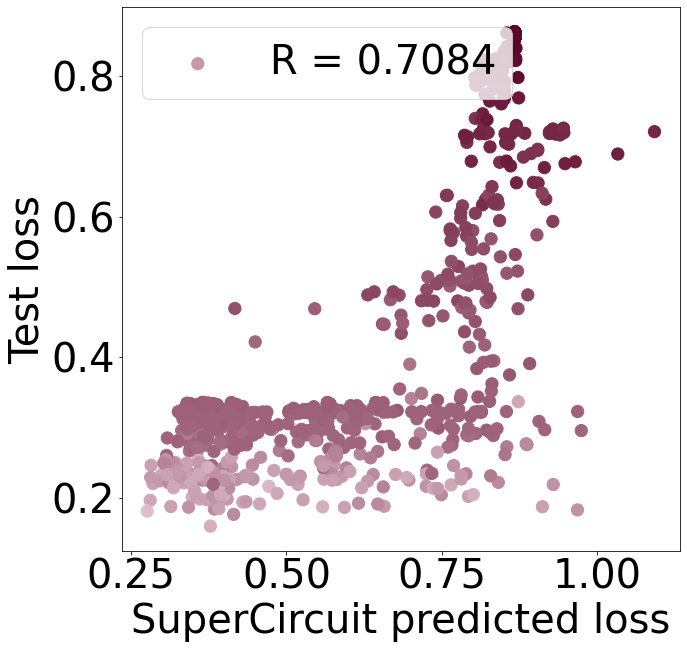}
        \includegraphics[width=0.49\textwidth]{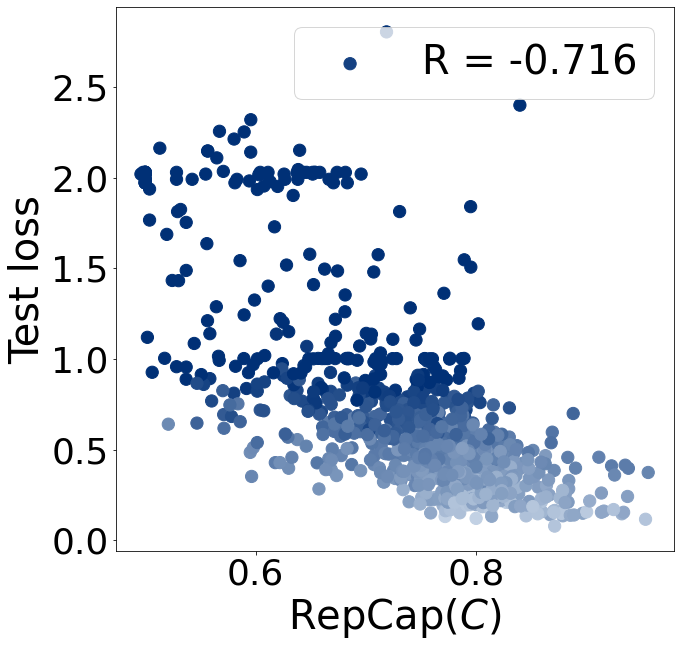}
        \caption{}
        \label{fig:rep_cor1}   
    \end{subfigure}
     \hfill
    \vspace{-0.1cm}
    \caption{\new{(a) \Metricone{} measures the degree of intra-class similarity and inter-class separation of the quantum states created by a circuit. Circuit 1 has a higher \metricone{}, and so is likely to perform better than Circuit 2. (b) Predicting circuit performance on the FMNIST-2 benchmark. \MO{} predicts circuit performance as well as a SuperCircuit, despite not involving gradient computation.}}
    \label{fig:correlations}
\end{figure*}

In comparison, SuperCircuit-based QCS works conflate the evaluation of circuit noise robustness and circuit performance, making the identification of low-fidelity circuits unnecessarily expensive. In these methods, the performance of a candidate circuits is evaluated using a validation data set $\mathcal{X_{\text{valid}}}$ on the target device. This requires executing $|\mathcal{X_{\text{valid}}}|$ circuits to evaluate every candidate circuit. This cost is often significantly larger than the constant number of circuits required to compute \MT{}, making identifying low performing circuits needlessly costly in SuperCircuit-based frameworks.


Figure~\ref{fig:clifford_correlation_realdevice} and~\ref{fig:clifford_correlation_simulator} show the correlation between the \MT{} and the fidelity of circuits executed on IBMQ-Guadalupe and IBMQ-Kolkata, and a noise model of Rigetti Aspen-M-2. In all cases, \MT{} is highly predictive of circuit fidelity.

\begin{hintbox}{\color{white}\textbf{Insight 3: Early rejection of low-fidelity circuits}}
Noise robustness evaluation is simpler than performance evaluation. Evaluating noise robustness first using \metrictwo{} enables early rejection of low-fidelity circuits, reducing evaluation costs.
\end{hintbox}
\begin{figure}[t]
    \centering
    \includegraphics[width=0.45\columnwidth]{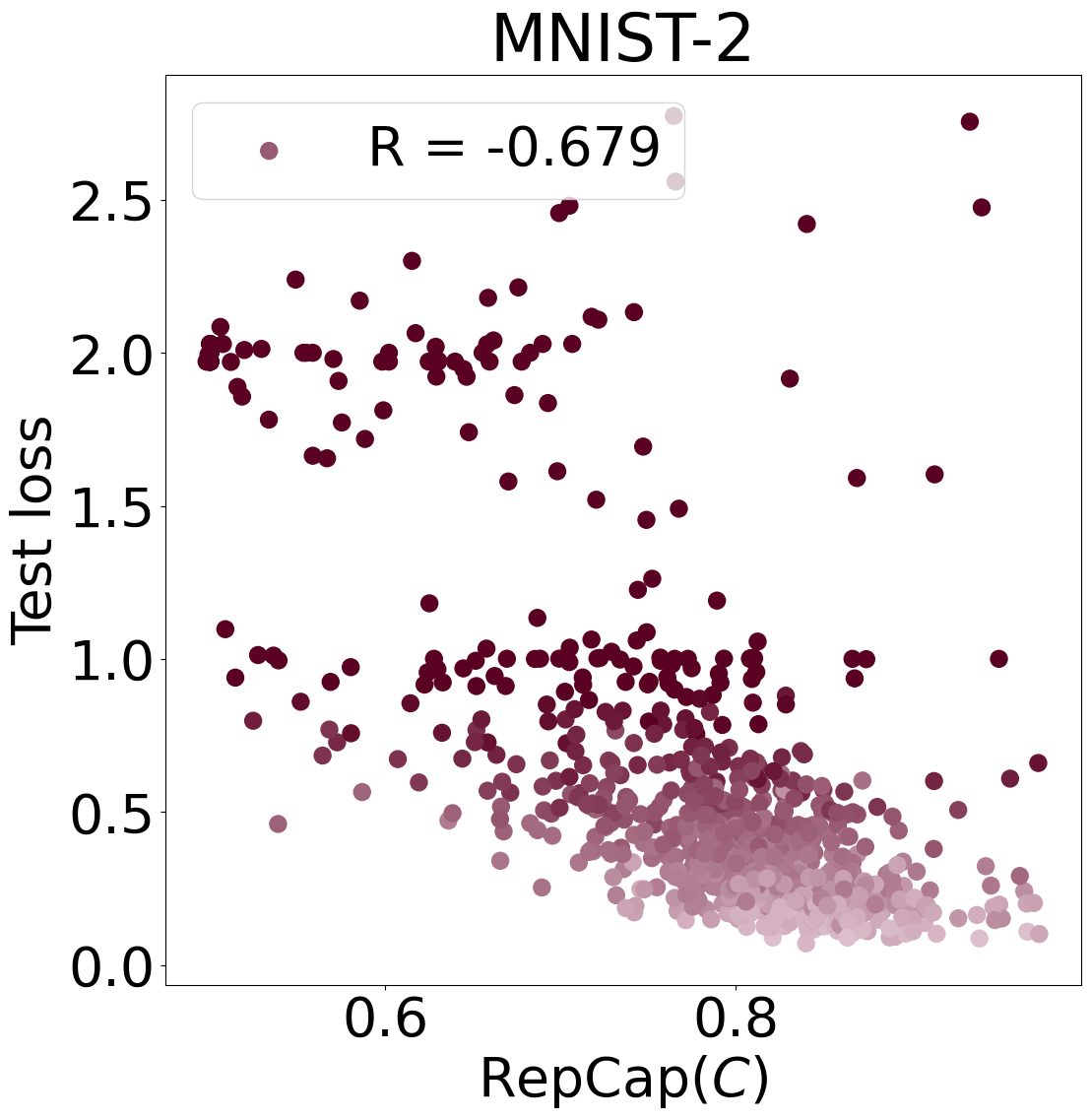}
    \hfill
    \includegraphics[width=0.45\columnwidth]{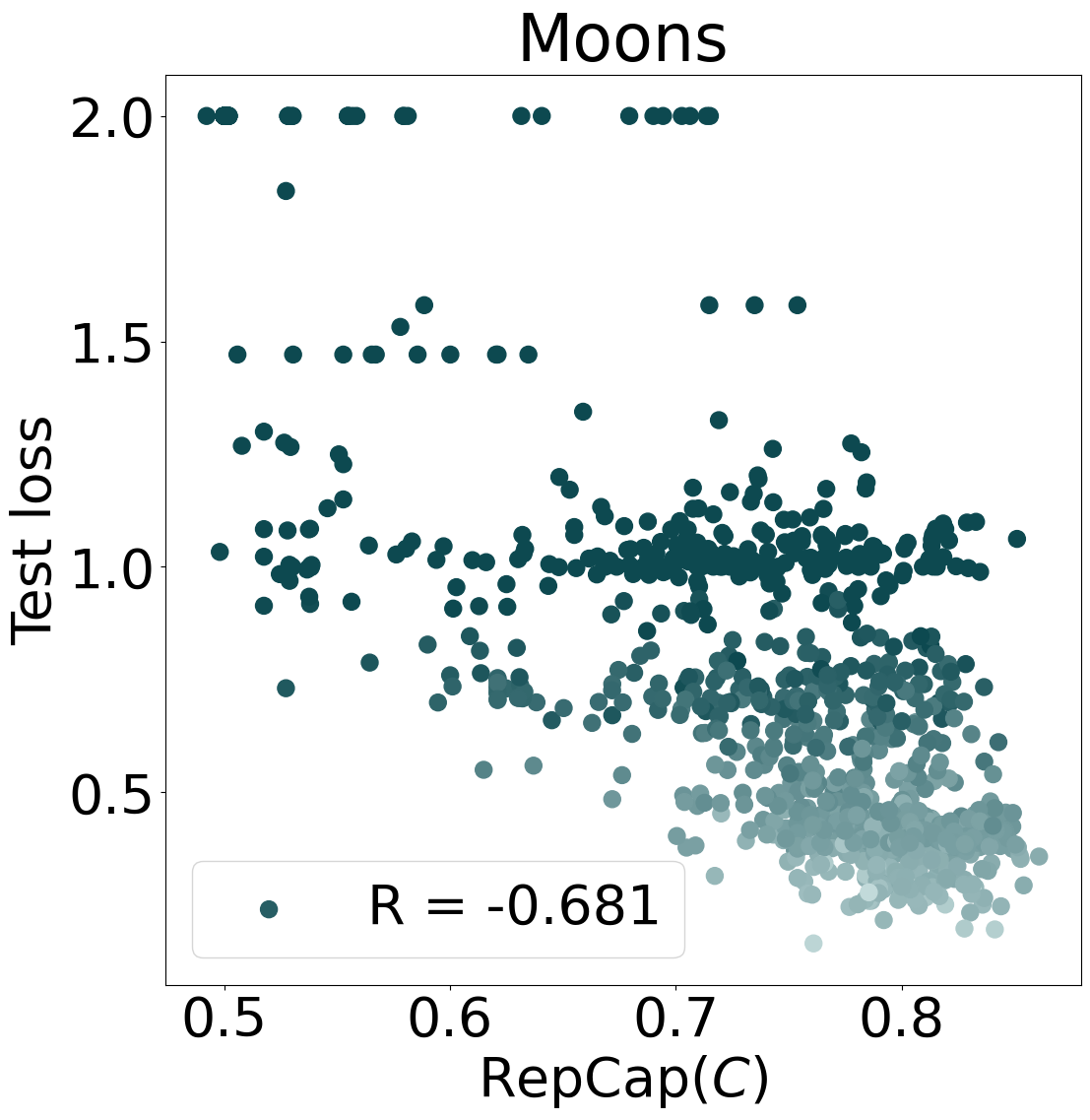}
    \hfill
    \vspace{-0.1cm}
    \caption{\Metricone{} is a strong predictor of performance for various QML tasks.}
    \label{fig:more_repcap}
\end{figure}

\section{Circuit performance evaluation}
\label{sec:metrics_explanations}

 Our calculations indicate that over 90\% of the circuit executions for SuperCircuit-based QCS methods are performed during SuperCircuit training, due to the high cost of computing gradients on quantum hardware via the parameter-shift rule. Thus, eliminating the initial training phase of QCS is crucial for resource efficiency.


\framework{} addresses this by introducing a novel, cheap-to-compute circuit performance predictor, \textit{\metricone{}} (\MO{}). \MO{} requires far fewer circuit executions than training-based evaluation strategies, but is just as effective at predicting circuit performance. \cref{fig:correlations} shows the correlations of SuperCircuit predicted losses and \MO{} with trained circuit performance on the FMNIST-2 task. Despite not requiring any training, \MO{} is as strong a predictor of performance as a trained SuperCircuit. \MO{} is a strong predictor of circuit performance on other QML benchmarks as well, as shown in \cref{fig:more_repcap}, and has a Spearman R correlation of 0.632 with circuit performance over all the QML benchmarks used for evaluation.

First, we define \MO{} and provide some intuition about it, and then elaborate on how to compute \MO{}. Then, we discuss how we combine \MT{} and \MO{} to rank the circuits remaining after the noise-guided circuit elimination process.


\subsection{Predicting circuit performance using \MO{}}
Intuitively, \MO{} measures the intra-class similarities and inter-class separation of the final quantum states for a variational circuit. As illustrated in~\cref{fig:rep_cap_diag}, the larger the inter-cluster distance and the smaller the intra-cluster distance in the final qubit state space, the higher the \MO{} of a variational circuit is. A higher \MO{} indicates that the circuit has a larger ``capacity" for representing the input data, and likely will perform better. 

\MO{} for circuit $C$ is defined as:

\begin{equation}
    \textrm{\MO{}}(C) = 1 - \frac{\| R_C - R_{\textrm{ref}}\|^{2}_{2}}{2 \cdot n_{c} \cdot d_{c}^2},
    \label{eq:repcap}
\end{equation}
where $d_c$ and $n_c$ are the number of samples selected from each class and the number of classes in $\mathcal{X}_{\text{train}}$, respectively. $||\cdot||_2$ is the Frobenius norm. $R_\text{ref}$ is a $d\times d$ reference matrix that captures the ideal circuit behaviour, i.e. $R_\text{ref}(i, j) = 1$ iff $y_i = y_j$, and $R_\text{ref}(i, j) = 0$ otherwise. $R_C$ is a $d\times d$  matrix that captures pairwise similarities between the representations of data points $x_i$ and $x_j$ created by circuit $C$, with $R_C{(i, j)} \in [0, 1]$. 

The entries of $R_C$ used in~\cref{eq:repcap} are the \textit{induced similarity} $\textrm{IS}_C$ between data points $x_i$ and $x_j$, i.e., 
\begin{equation}
    R_C{(i, j)} = \textrm{IS}_C(x_i, x_j).
\end{equation}

$\textrm{IS}_C(x_i, x_j)$ is defined as the averaged similarity of the output quantum states $\rho_C(x_i, \theta)$ and $\rho_C(x_j, \theta)$ of circuit $C$ when fed input data $x_i$ and $x_j$ over random parameters $\theta$. In practice, we use classical approximations $\hat{\rho}_C(x_i, \theta)$, computed via~\cref{alg:computesimilarities}, instead of states $\rho_C(x_i, \theta)$ in order to avoid executing circuits for every pair of samples $x_i$ and $x_j$:

\begin{equation}
    \textrm{IS}_C(x_i, x_j) = \frac{1}{n} \sum_{\theta = \theta_1}^{{\theta}_n} \textrm{Sim}(\hat{\rho}_C(x_i, \theta), \hat{\rho}_C(x_j, \theta))
\end{equation}

\begin{algorithm}[t]
\SetAlgoLined
    \KwIn{Circuit $C$; Input data $x$; variational parameters $\theta$; array of random rotation angles $\boldsymbol{\alpha}$ of size $n_{\text{bases}} \times n_{\text{meas}} \times 3$.}
    \KwOut{Classical approximation $\hat{\rho}_C(x, \theta)$ of $\rho_C(x, \theta)$}

    Append a set of $n_{\text{meas}}$ U3 gates to $C$ acting on the $n_{\text{meas}}$ qubits being measured\;
 
    Initialize $\hat{\rho}_C(x, \theta)$ as an empty list of length $n_{\text{bases}}$\;
 
    \For{$i = 1$ to $n_{\text{bases}}$}{
        Execute circuit $C(x, [\theta, \boldsymbol{\alpha}_i])$\;
        Construct probability distribution $P_{(\boldsymbol{\alpha}, i)}(x, \theta)$ using the outputs of $C(x, [\theta, \boldsymbol{\alpha}_i])$\;
        Set $\hat{\rho}_C(x, \theta)_i \leftarrow P$
    }
    
    \Return $\hat{\rho}_C(x, \theta)$;
    \caption{Constructing the classical approximation $\hat{\rho}_C(x, \theta)$ of a representation $\rho_C(x, \theta)$}
    \label{alg:computesimilarities}
\end{algorithm}
\vspace*{-0.2cm}

The angles $\{\theta_i\}$ can be uniformly sampled. Then the \textit{similarity} $\textrm{Sim}(x_i, x_j)$ between $x_i, x_j$ can be defined as the one minus Total Variational Distance (TVD) between their output states by $C$ using a randomized measurement protocol \cite{haug2021largescaleqml, huang2020classicalshadows, huang2022randomizedmeasurementtoolbox},\vspace*{-0.2cm} 
\begin{multline}
    \textrm{Sim}(\hat{\rho}_C(x_i, \theta), \hat{\rho}_C(x_j, \theta)) = \frac{1}{n_{\textrm{bases}}} \\ \sum_{k = 1}^{n_\textrm{bases}} 1 - \text{TVD}(\hat{\rho}_C(x_i, \theta)_k, \hat{\rho}_C(x_j, \theta)_k).    
\end{multline}
where $n_\textrm{bases}$ is the number of random bases we use to approximate the two states.

Computing \MO{} requires only $n_c \cdot d_c \cdot n_p$ circuit executions, where $n_p$ is the number of parameter initializations we average over. In practice, we take $d_c = 16$, $n_p = 32$. For $N_C$ circuits, \framework{} then requires $512 \cdot N_C \cdot n_c$ circuit executions. In contrast, SuperCircuit-based methods require $2 \cdot t \cdot |\mathcal{X_{\text{train}}}| \cdot \bar{p} + N \cdot |\mathcal{X_{\text{valid}}}|$ circuit executions to train a SuperCircuit for $t$ epochs via sampling subcircuits with $\bar{p}$ parameters each on average, and then evaluate the performance of $N$ subcircuits. The initial training overhead of a SuperCircuit scales very poorly with problem size due to the high cost of gradient computation, increasing the speedup of \framework{} with problem size. For example, \framework{} is only 44$\times$ faster than QuantumNAS on the 16-parameter Moons task (\cref{sec:experimental_setup}); for the 72-parameter MNIST-10 task, the speedup is over 5000$\times$. 
\ignore{
\renewcommand\theadalign{bc}
\renewcommand\theadfont{\bfseries}
\renewcommand\theadgape{\Gape[6pt]}
\renewcommand\cellgape{\Gape[6pt]}
\begin{table*}[h]
  \centering
    \begin{tabular}{ | c || c | c | c | c | c | c |}
        \hline
         \textbf{Dataset} & Moons-300 & Bank & Vowel-2 & Vowel-4 & MNIST-4 & FMNIST-4\\ 
         \hline
         \textbf{Spearman R Correlation} & -0.663 & -0.582 & -0.655 & -0.573 & -0.568 & -0.621\\
         \hline
    \end{tabular}
    \caption{The correlations between \metricone{} and trained loss computed for 1000 randomly sampled circuits each.}
    \label{tab:correlations_rep_cap}
\end{table*}
}



\begin{hintbox}{\color{white}\textbf{Insight 4: Predict circuit performance without training}}
By analyzing the output states of a circuit for different input data, we can cheaply estimate circuit performance as well as prior work.
\end{hintbox}

\subsection{Final evaluation of remaining circuits}
\label{sec:combining_metrics}
\framework{} combines \MT{}$(C)$ and \MO{}$(C)$ to accurately predict performance of the remaining circuits on the target device. The composite score $\text{Score}(C)$ for circuit $C$ is given by
\begin{equation}
    \text{Score}(C) = \text{\MT{}}(C)^{\alpha_{\text{CNR}}} \times \text{\MO{}}(C).
\end{equation}
\new{$\alpha_{\text{CNR}}$ is a hyperparameter controlling the relative importance of \MT{}$(C)$, with \MO{}$(C)$ and \MT{}$(C)$ being equally important at $\alpha_{\text{CNR}} = 1$. For all experiments, we set $\alpha_{\text{CNR}} = 0.5$. Finally, \framework{} returns the circuit with the highest $\text{Score}(C)$.}

\section{Experimental Setup}
\label{sec:experimental_setup}

\subsection{Benchmarks}

We conduct experiments on 9 different QML benchmarks. A summary of the benchmarks is shown in~\cref{tab:datasets}. Train and Test in~\cref{tab:datasets} refer to the number of samples used for training and testing, respectively. Params is the number of parameters in the circuits we search for in \cref{sec:evaluation}. Moons is a synthetic dataset generated using scikit-learn. We select a balanced subset of the original Bank dataset \cite{banknote_authentication_267}. We use classes $\{\textrm{0, 1}\}$ from the MNIST dataset for MNIST-2, $\{\textrm{0, 1, 4, 8}\}$ for MNIST-4, $\{\textrm{T-shirt, Trouser}\}$ from the FMNIST dataset for FMNIST-2, and $\{\textrm{T-shirt, Trouser, Bag, Ankle Boot}\}$ for FMNIST-4. For the MNIST and FMNIST benchmarks, we center crop the original images to 24x24 and downsample them to 4x4 using mean pooling, except for MNIST-10, which is downsampled to 6x6. 

The Vowel-2 and Vowel-4 datasets are constructed by merging 8 out of the 9 classes in the original Vowel dataset, and selecting the 10 most significant PCA dimensions.

\renewcommand\theadalign{bc}
\renewcommand\theadfont{\bfseries}
\renewcommand\theadgape{\Gape[6pt]}
\renewcommand\cellgape{\Gape[6pt]}
\begin{table}[t]
\vspace{-0.05in}
    \caption{Summary of the 9 QML benchmarks used for evaluations.}
  \centering
\setlength{\tabcolsep}{0.153cm} 
\renewcommand{\arraystretch}{1.1}
    \begin{tabular}{ l  c  c  c  c c}
        \toprule
         Benchmark & Classes & Data Dim. & Train & Test & Params\\
         \midrule
         Moons & 2 & 2 & \multicolumn{1}{r}{0.6K} & \multicolumn{1}{r}{0.12K} & 16\\
         Bank & 2 & 4 & \multicolumn{1}{r}{1.1K} & \multicolumn{1}{r}{0.12K} & 20\\ 
         MNIST-2 & 2 & 4x4 & \multicolumn{1}{r}{1.6K} & \multicolumn{1}{r}{0.4K} & 20\\
         MNIST-4 & 4 & 4x4 & \multicolumn{1}{r}{8K} & \multicolumn{1}{r}{2K} & 40\\
         FMNIST-2 & 2 & 4x4 & \multicolumn{1}{r}{1.6K} & \multicolumn{1}{r}{0.2K} & 32\\
         FMNIST-4 & 4 & 4x4 & \multicolumn{1}{r}{8K} & \multicolumn{1}{r}{2K} & 24\\
         Vowel-2 & 2 & 10 & \multicolumn{1}{r}{0.6K} & \multicolumn{1}{r}{0.12K} & 32\\
         Vowel-4 & 4 & 10 & \multicolumn{1}{r}{0.6K} & \multicolumn{1}{r}{0.12K} & 40\\ 
         MNIST-10 & 10 & 6x6 & \multicolumn{1}{r}{60K} & \multicolumn{1}{r}{10K} & 72\\ 
         \bottomrule
    \end{tabular}
    \label{tab:datasets}
\end{table}

\renewcommand\theadalign{bc}
\renewcommand\theadfont{\bfseries}
\renewcommand\theadgape{\Gape[6pt]}
\renewcommand\cellgape{\Gape[6pt]}
\begin{table}[t]
\vspace{-0.05in}
    \caption{Summary of the \emph{\textbf{real quantum hardware used}}.}
  \centering
\setlength{\tabcolsep}{0.13cm} 
\renewcommand{\arraystretch}{1.1}
    \begin{tabular}{l c c c c c}
        \toprule
         & & &\multicolumn{3}{c}{Median Error Rate $\downarrow$}\\
         \midrule
         Device & Qubits & QV & Readout & 1Q & 2Q\\
         \midrule
         OQC Lucy & 8 & -- & 1.3e-1 & 6.2e-4 & 4.4e-2\\
         Rigetti Aspen M-3 & 79 & -- & 8.0e-2 & 1.5e-3 & 9.3e-2\\
         IBMQ Jakarta & 7 & 32 & 2.6e-2 & 2.2e-4 & 8.5e-3\\
         IBM Nairobi & 7 & 32 & 2.4e-2 & 2.7e-4 & 9.6e-3\\
         IBM Lagos & 7 & 32 & 1.9e-2 & 2.1e-4 & 9.8e-3\\
         IBM Perth & 7 & 32 & 2.8e-2 & 2.8e-4 & 8.7e-3\\
         IBM Geneva & 16 & 32 & 2.7e-2 & 2.2e-4 & 1.1e-2\\
         IBM Guadalupe & 16 & 32 & 2.0e-2 & 2.9e-4 & 8.9e-3\\
         IBMQ Kolkata & 27 & 128 & 1.2e-2 & 2.3e-4 & 9.0e-3\\
         IBMQ Mumbai & 27 & 128 & 1.9e-2 & 2.0e-4 & 9.6e-3\\
         IBM Kyoto & 127 & -- & 1.4e-2 & 2.5e-4 & 9.1e-3\\
         IBM Osaka & 127 & -- & 1.7e-2 & 2.2e-4 & 1.0e-2\\
         \bottomrule
    \end{tabular}
    \label{tab:devices}
\end{table}

\begin{figure*}[t]
    \centering
    \begin{subfigure}{1\textwidth}
        \centering
        \includegraphics[width=1\textwidth]{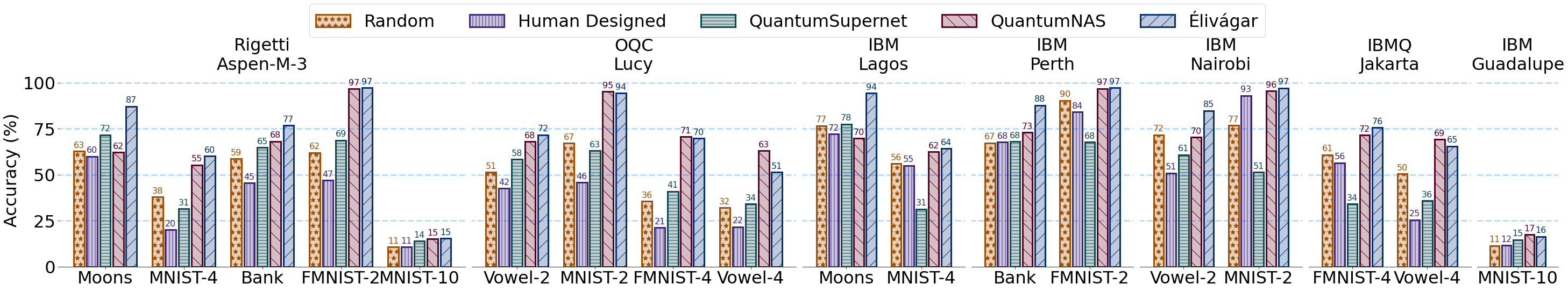}
        \caption{}
    \label{fig:noisy_sim_results}
     \end{subfigure}

    \begin{subfigure}{1\textwidth}
        \centering
        \includegraphics[width=1\textwidth]{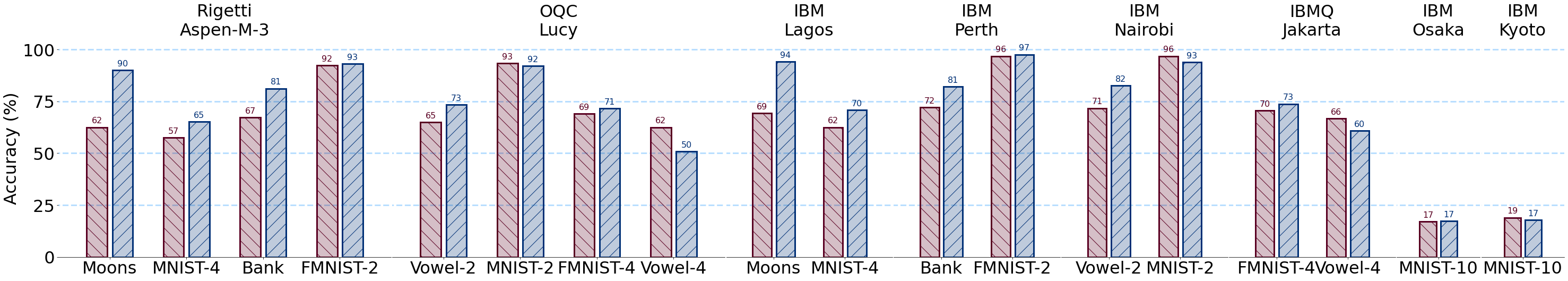}
        \caption{}
    \label{fig:real_dev_results}
    \end{subfigure}

    \caption{Results using (a) \emph{\textbf{noisy simulations}} with different noise models (each bar shows the mean of 25 runs), and (b) \emph{\textbf{real quantum hardware}} accessed via Amazon Braket and IBM Quantum. All plots show absolute classification accuracy \emph{(higher is better)}.}
     
    \label{fig:main_results}
\end{figure*}

\subsection{Backends and compiler configurations} 


The hardware devices used are listed in \cref{tab:devices}. We also use multiple noisy simulators use noise models based on IBM Nairobi, IBM Lagos, IBM Perth, IBMQ Jakarta, IBM Guadalupe, Rigetti Aspen-M-2, Rigetti Aspen-M-3, and OQC Lucy in order to perform evaluations, compute circuit fidelity and \MT{}. To compute \MO{} and perform circuit training, we use noiseless simulators from Pennylane \cite{bergholm2018pennylane} and TorchQuantum \cite{wang2021quantumnas}. Before running circuits on any hardware device or noisy simulator, we compile the circuit using the default Qiskit compiler with optimization level set to 3 for all competing methods except for QuantumNAS, for which we use level 2, and \framework{}, for which we use level 0.



\subsection{Training methodology}
All circuits are trained using the same methodology to ensure fair comparisons. We train circuits for 200 epochs using a batch size of 128 and optimize parameters using the Adam optimizer with learning rate 0.01. No weight decay or learning rate scheduling is used. The training objective is to minimize the classification loss between circuit predictions and the labels of the training set. We perform training on noiseless simulators using the TorchQuantum package \cite{wang2021quantumnas}, and use an AWS ml.g4dn.xlarge instance with 16GB of memory and an NVIDIA T4 GPU.

\subsection{Competing methods}

We compare \framework{} to four competing methods. Every circuit compared to, including those chosen by \framework{}, all contain the same number of parameters in order to ensure a fair comparison. The number of parameters used in the circuits for each benchmark is given in \cref{tab:datasets}.

\textbf{Random:} we randomly generate 25 circuits using the RXYZ + CZ gateset from \cite{wang2021quantumnas}, and report the average performance of the circuits.

\textbf{Human designed (baseline):} We use three data embedding schemes (angle, amplitude, and IQP embedding) paired with paired with the commonly used BasicEntanglerLayers template from the Pennylane\cite{bergholm2018pennylane} library. We train all three circuits and report the average performance. 

\textbf{QuantumSupernet \cite{dacheng2020quantumsupernet}:} We modify the SuperCircuit training procedure proposed in~\cite{dacheng2020quantumsupernet} and use mini-batch gradient descent with batch size 32 instead of full-batch gradient descent to ensure all SuperCircuit parameters are updated sufficiently. All other hyperparameters are taken from \cite{dacheng2020quantumsupernet}.

\textbf{QuantumNAS \cite{wang2021quantumnas}:} We use the same hyperparameters for SuperCircuit training and evolutionary search as \cite{wang2021quantumnas}, and use the RXYZ + CZ gateset since it performs the best out of the 6 gatesets in~\cite{wang2021quantumnas} before pruning. We do not perform iterative pruning for trained QuantumNAS circuits as the pruning phase does not contribute towards circuit search. 

\subsection{Hyperparameters for \framework{}}
\label{sec:elivagar-hyperparams}
We randomly sample $d_c = 16$ data points from each class of $\mathcal{X}_{\textrm{train}}$ and $n_p = 32$ parameter initializations to compute \MO{}. We use $M = 32$ Clifford replicas to compute \MT{}, and reject circuits with \MT{} scores less than 0.7 or outside the top 50\%. We construct composite scores using $\alpha_{\text{CNR}} = 0.5$. To mitigate the effects of random circuit sampling, we repeat our workflow 25 times, and report the average performance.  


\subsection{Figure of merit}
\label{sec:fom}



We measure the performance of a circuit using the classification accuracy it obtains over a dedicated test set for each benchmark. We highlight the difference between circuit accuracy and circuit fidelity in this work: the classification accuracy of a circuit is the fraction of predictions made by the circuit that are correct, i.e. match the ground truth label of the associated sample for which the prediction was made; while fidelity measures the degree to which the outputs of a quantum circuit remain unaffected by hardware noise.
\section{Results}
\label{sec:evaluation}

\subsection{Performance on QML benchmarks}
\cref{fig:noisy_sim_results} shows performance on 9 different QML benchmarks, using \emph{\textbf{noisy simulators}} with noise models of Rigetti Aspen-M-3, OQC Lucy,  IBM Lagos, IBM Perth, IBM Nairobi, IBMQ Jakarta. \cref{fig:real_dev_results} shows the results of evaluations performed on \emph{\textbf{real quantum hardware}}. \framework{} is competitive with or outperforms QuantumNAS on all benchmarks except Vowel-4. Circuit performance on the Rigetti Aspen-M-3 and OQC Lucy devices is worse than on the IBM devices due to their higher noise levels (see \cref{tab:devices}). On average, \framework{} achieves \perfgain{} higher accuracy than QuantumNAS, and \perfgainhuman{} higher accuracy than the human designed baseline.

\begin{table}[t]
  \centering
\setlength{\tabcolsep}{0.1cm} 
\renewcommand{\arraystretch}{1.1}
\begin{center}
\caption{Absolute runtimes (in minutes) for \framework{} and QuantumNAS for different benchmarks when using \emph{\textbf{classical simulators}}, and speedups experienced by \framework{} when running on \emph{\textbf{classical simulators}} (C) and \emph{\textbf{real quantum hardware}} (Q).}
    \label{tab:resources}
    \begin{tabular}{l c c c c}
    \toprule
    Benchmark & QNAS & \framework{} & Speedup (C) & \textbf{Speedup (Q)}\\
    \midrule
    Moons & \multicolumn{1}{r}{43.4} & \multicolumn{1}{r}{7.7} & \multicolumn{1}{r}{5.6x} & \multicolumn{1}{r}{\textbf{44x}}\\
    Vowel-4 & \multicolumn{1}{r}{63.3} & \multicolumn{1}{r}{8.9} & \multicolumn{1}{r}{7.0x} & \multicolumn{1}{r}{\textbf{77x}}\\
    Vowel-2 & \multicolumn{1}{r}{58.6} & \multicolumn{1}{r}{9.4} & \multicolumn{1}{r}{6.2x} & \multicolumn{1}{r}{\textbf{104x}}\\
    Bank & \multicolumn{1}{r}{48.9} & \multicolumn{1}{r}{7.6} & \multicolumn{1}{r}{6.4x} & \multicolumn{1}{r}{\textbf{119x}}\\
    MNIST-2 & \multicolumn{1}{r}{143.6} & \multicolumn{1}{r}{7.8} & \multicolumn{1}{r}{18.6x} & \multicolumn{1}{r}{\textbf{182x}}\\
    FMNIST-2 & \multicolumn{1}{r}{184.3} & \multicolumn{1}{r}{8.3} & \multicolumn{1}{r}{22.0x} & \multicolumn{1}{r}{\textbf{282x}}\\
    FMNIST-4 & \multicolumn{1}{r}{228.6} & \multicolumn{1}{r}{11.1} & \multicolumn{1}{r}{20.7x} & \multicolumn{1}{r}{\textbf{646x}}\\
    MNIST-4 & \multicolumn{1}{r}{174.2} & \multicolumn{1}{r}{15.4} & \multicolumn{1}{r}{11.3x} & \multicolumn{1}{r}{\textbf{1046x}}\\
    MNIST-10 & \multicolumn{1}{r}{1007.8} & \multicolumn{1}{r}{35.5} & \multicolumn{1}{r}{28.4x} & \multicolumn{1}{r}{\textbf{5220x}}\\
    \midrule
    \textbf{GMean} & \multicolumn{1}{r}{} & \multicolumn{1}{r}{} & \multicolumn{1}{r}{\textbf{11.7x}} & \multicolumn{1}{r}{\textbf{271x}}\\
    \bottomrule
    \end{tabular}
    \end{center}
    \vspace*{-0.5cm}
\end{table}

\subsection{Runtime speedups}

We consider two hardware setups when measuring the speedup of \framework{} over QuantumNAS, which we elaborate on below.

\subsubsection{Using classical simulators}
\label{sec:classical_runtime_analysis}

Using classical simulators results in reduced training costs as gradients can be computed efficiently using backpropagation. Additionally, training and inference on noiseless simulators can be performed in a batched manner, further speeding up the QCS process. We provide absolute runtime values for both QuantumNAS and \framework{} in \cref{tab:resources}. Even when using classical simulators and backpropagation, which disproportionately benefits training-heavy frameworks such as QuantumNAS, \framework{} is 11.7$\times$ faster than QuantumNAS on average. Moreover, the speedup achieved by \framework{} increases with benchmark size despite the reduced cost of gradient computation, highlighting \framework{}'s resource-efficiency.

\subsubsection{Using quantum hardware}
\label{sec:classical_runtime_analysis}

In this scenario, all training is done via the parameter-shift rule \cite{wierichs2022parametershift}, which results in drastically increased training costs, as explained in \cref{sec:background}. \new{Unfortunately, the high variance in the times required to run circuits on quantum hardware via a cloud computing platform} make it near impossible to reliably estimate runtimes via wall-clock time. Consequently, the most reliable way to estimate the speedup achieved by \framework{} is to compare the number of circuit executions required by both methods for each benchmark. \cref{tab:resources} shows the speedups achieved by \framework{}, which is \speedup{} times faster than QuantumNAS on average. We provide a detailed breakdown of this speedup in \cref{sec:source_of_speedup}.

\section{Performance Analysis}
\label{sec:perf_analysis}

\begin{table}[t]
  \centering
\setlength{\tabcolsep}{0.34cm} 
\renewcommand{\arraystretch}{1.1}
\begin{center}
    \caption{Characteristics of \framework{}-generated (no optimization) and device-unaware random circuits (SABRE~\cite{li2018sabre} + Qiskit compiler level 3) run on \textbf{\emph{real quantum hardware}}. Fidelity refers to the extent to which circuit outputs remain unaffected by noise, and is different from circuit accuracy (see \cref{sec:fom}).}
    \label{tab:device_aware_fids}
    \begin{tabular}{l c c c}
        \toprule
         Policy & 2Q gates & \makecell{2Q gates after\\compilation $\downarrow$} & Fidelity $\uparrow$\\
        \midrule
        \multicolumn{4}{c}{OQC Lucy}\\
        \midrule
        \imp{SABRE} & \imp{7.04} & \imp{16.22} & \imp{0.595}\\
        \imp{\textbf{\framework{}}} & \imp{\textbf{7.04}} & \imp{\textbf{7.04}} & \imp{\textbf{0.706}}\\
        \midrule
        \multicolumn{4}{c}{IBM-Geneva}\\
         \midrule
        \imp{SABRE} & \imp{6.48} & \imp{15.24} & \imp{0.615}\\
        \imp{\textbf{\framework{}}} & \imp{\textbf{6.48}} & \imp{\textbf{6.48}} & \imp{\textbf{0.714}}\\
        \midrule       
        \multicolumn{4}{c}{IBMQ-Kolkata}\\
        \midrule
        \imp{SABRE} & \imp{9.52} & \imp{22.16} & \imp{0.741}\\
        \imp{\textbf{\framework{}}} & \imp{\textbf{9.52}} & \imp{\textbf{9.52}} & \imp{\textbf{0.848}}\\
        \midrule
        \multicolumn{4}{c}{IBMQ-Mumbai}\\
        \midrule
        \imp{SABRE} & \imp{9.75} & \imp{25.00} & \imp{0.634}\\
        \imp{\textbf{\framework{}}} & \imp{\textbf{9.75}} & \imp{\textbf{9.75}} & \imp{\textbf{0.804}}\\
        \bottomrule
    \end{tabular}
    \end{center}
    \vspace*{-0.2cm}
\end{table}

\subsection{Device-aware circuit generation}
\label{sec:device_aware_gen}

To analyze \framework{}'s circuit generation strategy, we compare device-aware circuits generated via \cref{alg:candidategen} with randomly generated, device-unaware circuits. We generate pairs of device-aware and -unaware circuits with the same number of 1- and 2-qubit gates before compilation to ensure a fair comparison. Device-aware circuits are run unoptimized, but device-unaware circuits are optimized using level 3 of the Qiskit compiler and SABRE~\cite{li2018sabre}.\new{~\cref{tab:device_aware_fids} shows that device-aware circuits have 18.9\% higher fidelity on average. Thus, \framework{}'s circuit generation strategy not only eliminates the need for a circuit-mapping co-search, but also boosts circuit noise robustness, reducing performance degradation.}

\begin{table}[t]
  \centering
\setlength{\tabcolsep}{0.09cm} 
\renewcommand{\arraystretch}{1.1}
\begin{center}
\caption{Compiled circuit statistics for \framework{} and competing methods run on \emph{\textbf{real quantum hardware}}.}
    \label{tab:circ_stats}
    \begin{tabular}{l c c c c}
    \toprule
    Method & 1Q Gates $\downarrow$ & 2Q Gates $\downarrow$ & Depth $\downarrow$ & Acc. $\uparrow$\\
    \midrule
    \multicolumn{5}{c}{Vowel-2 (32 params.) on IBM Nairobi}\\
    \midrule
    Random & 171 & 52 & 139 & 0.653\\
    Human Designed & 194 & 52 & 139 & 0.508\\
    QuantumSupernet & 118 & 51 & 104 & 0.533\\
    QuantumNAS & 67 & 9 & 37 & 0.716\\
    \textbf{\framework{}} & \textbf{61} & \textbf{7} & \textbf{30} & \textbf{0.825}\\
    \midrule
    \multicolumn{5}{c}{MNIST-4 (40 params.) on IBM Lagos}\\
    \midrule
    Random & 201 & 60 & 163 & 0.552\\
    Human Designed & 252 & 155 & 260 & 0.550\\
    QuantumSupernet & 218 & 122 & 206 & 0.300\\
    QuantumNAS & 61 & 5 & 32 & 0.625\\
    \textbf{\framework{}} & \textbf{45} & \textbf{4} & \textbf{20} & \textbf{0.642}\\
    \midrule
    \multicolumn{5}{c}{MNIST-10 (72 params.) on IBM Osaka}\\
    \midrule
    Random & 89 & 87 & 211 & 0.128\\
    Human Designed & 355 & 228 & 344 & 0.113\\
    QuantumNAS & 121 & 22 & 55 & 0.190\\
    \textbf{\framework{}} & \textbf{107} & \textbf{24} & \textbf{49} & \textbf{0.178}\\
    \bottomrule
    \end{tabular}
    \end{center}
\end{table}

\subsection{Circuit statistics}
The statistics for the compiled and optimized circuits used for the Bank and Vowel-2 benchmarks are shown in \cref{tab:circ_stats}. Random and Human Design are noise and device unaware, resulting in large, deep circuits (depths 163, 260 for MNIST-4, and 211, 344 for MNIST-10) with many two-qubit gates even after optimization using level 3 of the Qiskit compiler. As a result, these circuits experience a large reduction in accuracy when run on real devices compared to their noise-free performance. QuantumSupernet is noise and device-aware, but uses a deep embedding subcircuit that requires multiple layers of entangling CRY gates, and thus undergoes significant accuracy reduction as well. The circuits found by QuantumNAS and \framework{} are much shallower due to the evolutionary search and \MT{} incentivizing the selection of shallow, noise-robust circuits, respectively.

However, despite being similarly noise-robust and shallow, the circuits found by \framework{} perform significantly better than those found by QuantumNAS, likely due to the difference in the embeddings used by the circuits. We explore the effect of searching for data embeddings in detail in \cref{sec:accuracy_breakdown}.

\begin{figure}[t]
    \centering
    \includegraphics[width=\columnwidth]{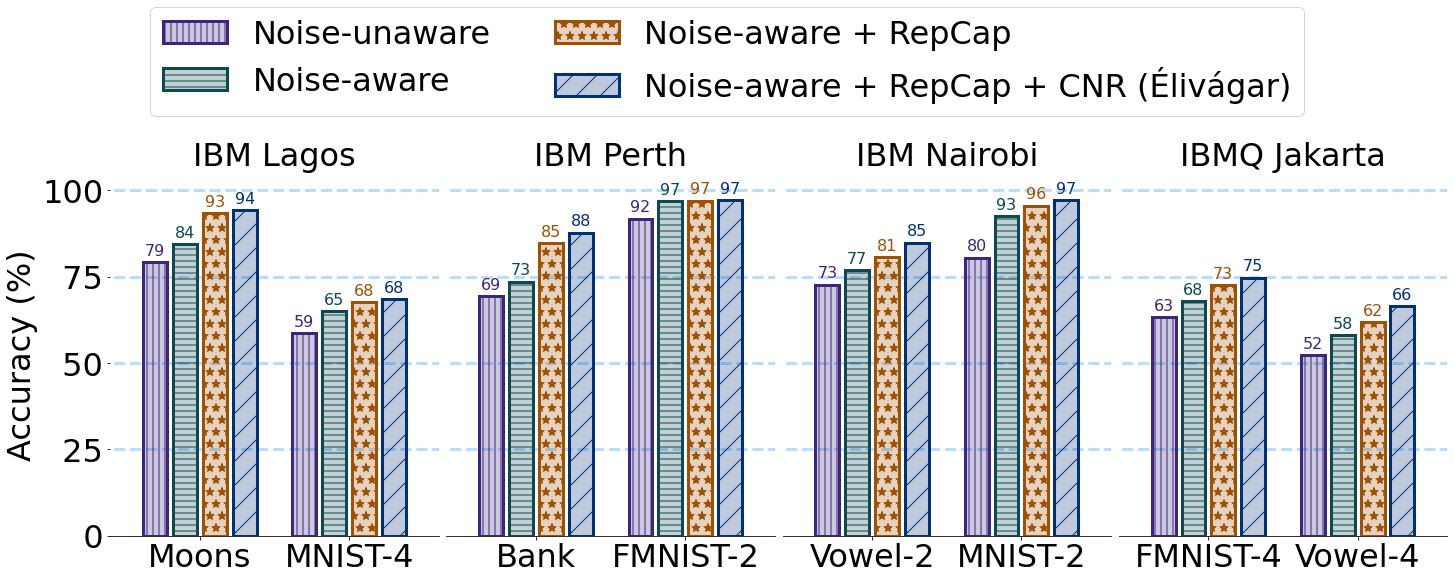}
    \caption{\new{Each component of \framework{} - device- and noise-aware circuit generation, \MO{}, and \MT{} - contributes to the  final accuracy of the selected circuit. All values are absolute classification accuracies obtained on \emph{\textbf{noisy simulators}}. Each bar shows the mean of 25 runs.}}
    \label{fig:accuracy_breakdown}
\end{figure}

\subsection{Breakdown of performance improvement}
\label{sec:accuracy_breakdown}
We break down \framework{}'s performance improvement into three parts: device-aware circuit generation (\cref{sec:circuit_generation}), \MO{} and data embeddings (\cref{sec:metrics_explanations}), and \MT{} (\cref{sec:clifford_noise}) in \cref{fig:accuracy_breakdown}. We use 8 benchmarks from \cref{tab:datasets}, and compare \framework{} to three baselines: (1) device- and noise-unaware circuits, (2) hardware-efficient device- and noise-aware random circuits generated via \cref{alg:candidategen}, and (3) circuits found by \framework{} using only \MO{}, i.e. without using \MT{} to rank circuits. 

Using device- and noise-aware circuits increases accuracy over device-unaware circuits by 5\%, further demonstrating the improved noise-robustness of circuits generated using \cref{alg:candidategen}. The biggest accuracy improvement (6\%) is obtained by using \MO{} to select circuits instead of choosing randomly, highlighting the importance of strong predictors of circuit performance. 

\begin{figure}[t]
    \centering
    \includegraphics[width=\columnwidth]{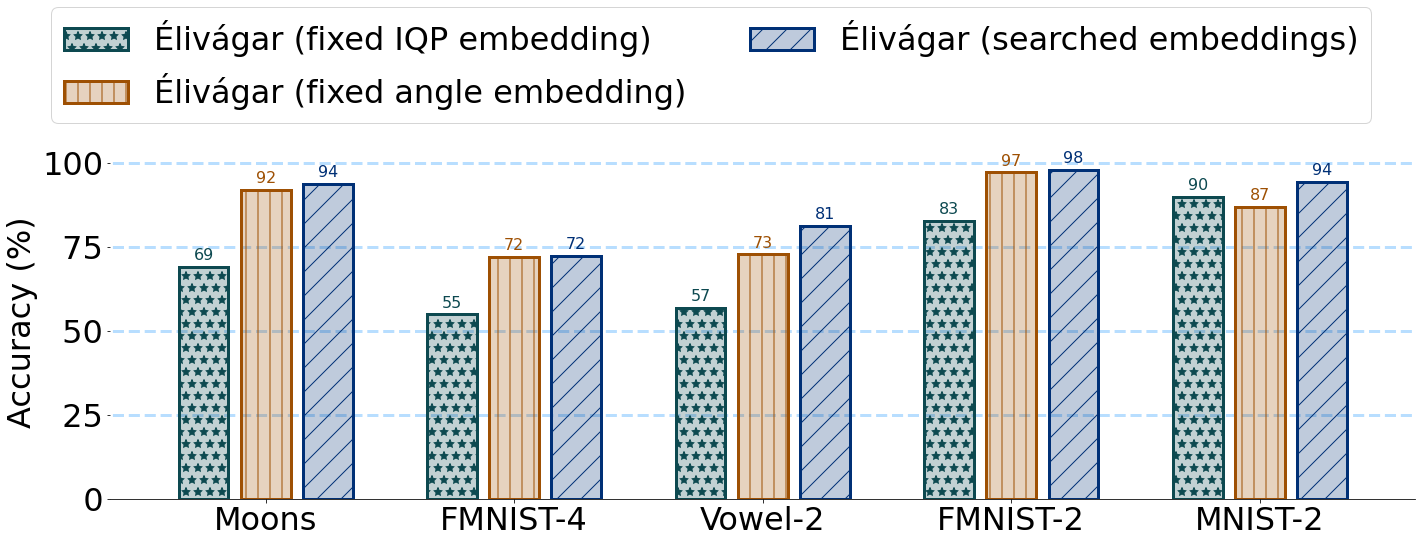}
    \caption{Performance of \framework{} when searching for data embeddings versus when using a fixed data embedding. All values are absolute classification accuracies obtained on \emph{\textbf{noiseless simulators}}. Each bar shows the mean of 25 runs.}
    \label{fig:fixed_vs_searched_embedding}
\end{figure}

Using RepCap to select circuits allows \framework{} to search for both data embeddings and variational gates. To isolate the effect of searching for data embeddings, we compare \framework{} with versions of \framework{} that use fixed angle~\cite{qml} and IQP embeddings~\cite{havlek2019suplearninginquantumenhancedfeaturespaces}. We evaluate the three versions of \framework{} using a noiseless simulator to eliminate the effects of noise. As shown in \cref{fig:fixed_vs_searched_embedding}, \framework{} obtains 5.5\%  and 20\% higher accuracy when searching for data embeddings than when using a fixed angle and IQP embedding, respectively. This is because searching for embeddings allows \framework{} to co-search for suitable combinations of embeddings and variational gates, allowing it to find higher performance circuits than when using a fixed embedding. Thus, almost all of the accuracy gains achieved by using \MO{} is due to the data embedding search, reinforcing the significance of data embeddings in determining performance on QML tasks.

Finally, using \MT{} in addition to \MO{} further increases accuracy by 2\% on average. Using both predictors results in better performance than only using \MO{} as while using \MO{} leads to \framework{} choosing circuits with high noiseless accuracy, \MO{} is not device- and noise-aware and thus may choose circuits that are not noise-robust, leading to large accuracy degradations when run on real hardware. Using both predictors allows \framework{} to balance circuit learning capability and expressivity with circuit noise robustness to find a circuit that can accurately learn the target dataset while also being robust to hardware noise.

\subsection{Breakdown of runtime speedup}
\label{sec:source_of_speedup}
When training on quantum hardware, the speedup of \framework{} over QuantumNAS comes from (1) using \metricone{} for performance evaluation, (2) early rejection of low-fidelity circuits, and (3) eliminating the circuit-mapping co-search by generating device-aware circuits. Using \metricone{} speeds up \framework{} by 16$\times-$78$\times$ for the benchmarks in~\cref{sec:evaluation}, growing with problem size. Since we use conservative CNR hyperparameter values, the speedup due to early rejection for all tasks is 2$\times$. Eliminating the circuit-mapping co-search provides a speedup of 1.4$\times-$33.4$\times$, growing with problem size.

\begin{figure}[t]
    \centering
    \begin{subfigure}{0.48\textwidth}
        \centering
        \includegraphics[width=1\columnwidth]{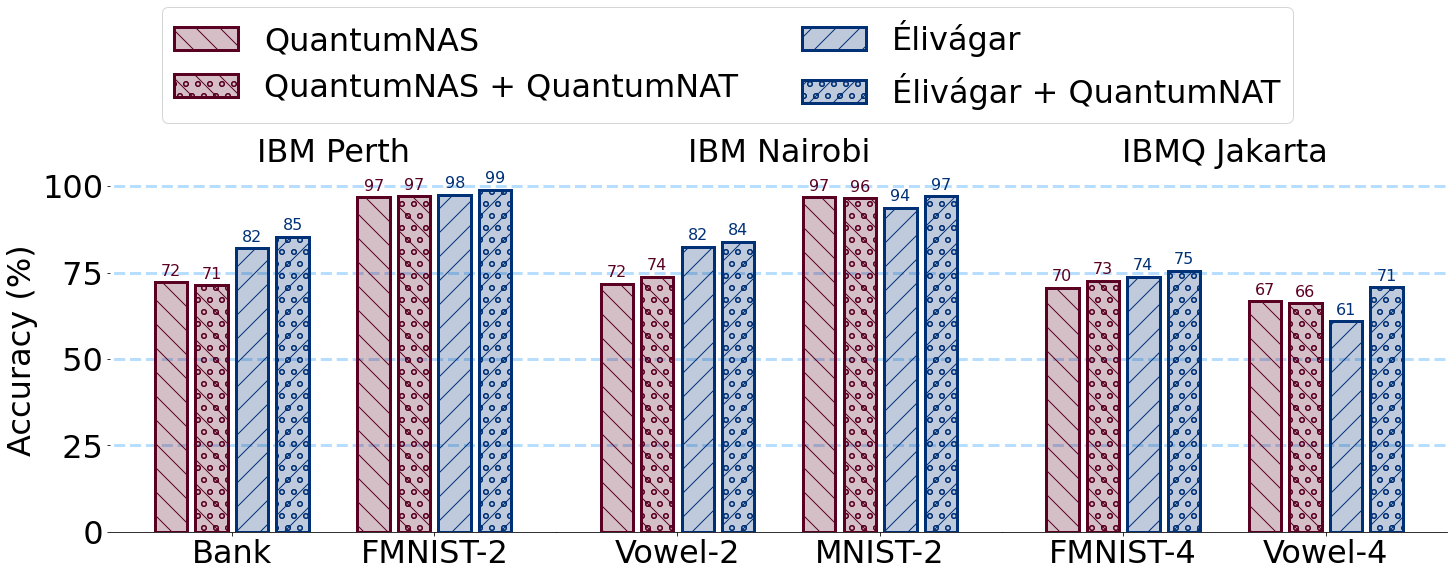}
        \caption{}
        \label{fig:quantumnat_comparison}
    \end{subfigure}
    \begin{subfigure}{0.48\textwidth}
        \centering
        \includegraphics[width=1\columnwidth]{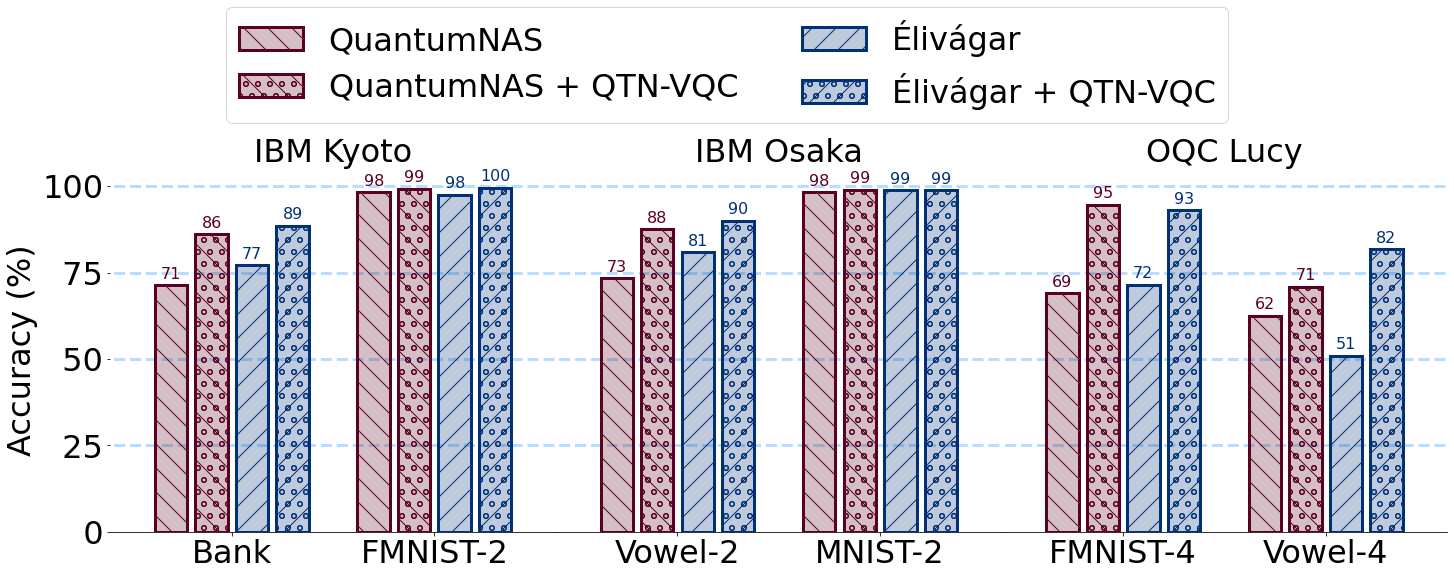}
        \caption{}
        \label{fig:qtn_vqc_comparison}
    \end{subfigure}    
    \caption{Results on \textbf{\emph{real quantum hardware}} when performing training and inference with and without (a) QuantumNAT \cite{wang2021quantumnat} and (b) QTN-VQC \cite{qi2021qtnvqc}. All values are absolute classification accuracies.}
\end{figure}

\subsection{Compatibility with other QML frameworks}

\framework{} is compatible with frameworks targeting the data preprocessing and training stages of the QML pipeline, such as QTN-VQC \cite{qi2021qtnvqc} and QuantumNAT \cite{wang2021quantumnat}, and makes no assumptions about how these tasks are performed. These frameworks can be combined with \framework{} to further boost circuit performance. \cref{fig:quantumnat_comparison} compares the performance of \framework{} and QuantumNAS with and without QuantumNAT \cite{wang2021quantumnat}, a framework that aims to increase the noise robustness of circuits during training and inference. \framework{} obtains 2.2\% higher accuracy than QuantumNAS + QuantumNAT, and 5.5\% higher accuracy when paired with QuantumNAT.

We further combine \framework{} and QuantumNAS with QTN-VQC, a framework that performs classical preprocessing of input data via a trainable tensor network, with results shown in \cref{fig:qtn_vqc_comparison}. Even when using QTN-VQC, which significantly boosts performance due to the added classical trainable parameters in the tensor network used for preprocessing, \framework{} outperforms QuantumNAS by 2.4\% on average.

\section{Related works}
\label{sec:discussion}
\subsection{Quantum Machine Learning}
Several theoretical works have illustrated potential quantum advantage in tasks such as classification~\cite{nature_communication_2021_power_of_data_qml, abbas2020power} and regression~\cite{schuld2022fourierinterpolation}. Various approaches to QML have been proposed, including using quantum circuits as kernel methods~\cite{schuld2021qmlmodelsarekernelmethods, havlek2019suplearninginquantumenhancedfeaturespaces, nature_communication_2021_power_of_data_qml, haug2021largescaleqml, glick2021covariantkernels, kubler2021inductivebias, tao2021noisykernels, liu2021robustandrigorousspeedup, canatar2022bandwidthinquantumkernels}, and as quantum neural networks~\cite{Silver_Patel_Tiwari_2022, optic, wang2021roqnn, coles2021overparametrizationinqnns, farhi2018classification, 2019PQC, 2019Expressibility, zoufal2021qbms, nguyen2022equivariantqml, qi2021qtnvqc}. Multiple works~\cite{2019Expressibility, abbas2020power} introduced metrics that estimate circuit performance, but they are unsuitable for QCS due to their high cost.

\subsection{Noise-aware quantum compilation and training}
Recent compilation works explore different circuit transformations to improve noise robustness, including gate cancellation~\cite{xu2022quartz, nam_optimization, efficient_exact_matching}, qubit mapping and routing~\cite{patel2022geyser, lye2015determining, oddi2018greedy, baker2020time, wille2014optimal, nannicini2021optimal, smith2021error, liu2022not, liu2021relaxed, noiseadaptive, tannu2019not}, error mitigation~\cite{jigsaw, adapt, disq, qraft}, circuit synthesis~\cite{patel2022quest, weiden2022topology, patel2022robust, younis2021qfast, peterson2022optimal}, and pulse-level optimizations~\cite{gokhale2020optimized,shi2019optimized, liang2022pan, meirom2022pansatz}. Some of these works~\cite{adapt, das2022imitationgame, ravi2022cliffordcanarycircuits} use Clifford replicas to characterize device noise or create enhanced compilation passes. \framework{}, in contrast, applies Clifford replicas to QML, and uses them to perform a noise-guided search for QML circuits. Other works~\cite{wang2021quantumnat, wang2022qoc} focus on increasing noise robustness for QML circuits through improved training procedures, an approach complementary to \framework{}, as these techniques can be combined with \framework{} to improve the training of circuits selected by \framework{}.



\subsection{QCS for Variational Quantum Algorithms (VQAs)}
A few studies have explored QCS for VQAs~\cite{vqa_survey}, such as QAOA~\cite{yao2022mctsqaoa}, VQE~\cite{huang2022qceat}, and state preparation~\cite{hong2020dqas, metaqas, continuousQAS}. These works largely adopt techniques from classical NAS. For example, QCEAT~\cite{huang2022qceat} is similar to evolutionary search-based NAS~\cite{quoc2018evosearchnas}, and MCTS-QAOA~\cite{yao2022mctsqaoa} uses a reinforcement learning-based search algorithm inspired by~\cite{baker2016rlnas, quoc2016rlnas}. These frameworks can be extended to perform QCS for QML tasks, despite being originally designed for VQAs. However, since they rely on classically-inspired search and candidate evaluation mechanisms, these methods face the same issues as QCS works for QML, and are extremely slow even for small QML tasks. We compare \framework{} with \cite{huang2022qceat, yao2022mctsqaoa} in \cref{tab:qcs_comparison}.

\subsection{Data embeddings}
Several works \cite{schuld2021encodingexpressivity, schuld2022fourierinterpolation, sweke2021encodingdependentbounds} have theoretically investigated the effect of data embeddings on circuit performance through the lens of generalization bounds and Fourier series. These works focus on developing theory that relates the data embeddings used to trained circuit performance. As such, they use simplified settings that do not account for hardware noise or limited device connectivity, and place strict constraints of the structures of the circuits used. Thus, they cannot be used as tools for QCS directly. In contrast, \framework{} solves the near-term problem of identifying performant circuits for QML applications on NISQ hardware, which is a fundamentally different task. \framework{}, however, applies these theoretical insights on data embeddings to QCS by searching for both optimal data embeddings and variational gates for a QML task, which allows \framework{} to outperform prior QCS works.

\subsection{QCS for QML}
 A few studies have focused on QCS for QML tasks, such as QuantumNAS~\cite{wang2021quantumnas} and QuantumSupernet~\cite{dacheng2020quantumsupernet}. These methods are both based on the classical Supernet \cite{pham2018enas} framework, and have been extensively discussed in \cref{sec:intro}, and compared with \framework{} in \cref{sec:evaluation}. \framework{} \new{eschews} the SuperCircuit that \cite{wang2021quantumnas, dacheng2020quantumsupernet} use in favor of a novel QCS pipeline that leverages training- and gradient-free circuit performance and noise robustness predictors to avoid expensive gradient computation. Thus, \framework{} avoids the runtime bottlenecks of SuperCircuit-based QCS frameworks, and is able to perform QCS with low overheads. In~\cref{tab:qcs_comparison}, we present a summary of the differences between \framework{} and prior works.

\section{Conclusion}
In this work, we present \framework{}, a noise-guided, resource-efficient QCS framework that searches for high-performance variational quantum circuits for QML tasks. \framework{} proposes that QCS methods should be tailored to QML tasks and carefully consider realistic  constraints of QML algorithms and NISQ quantum hardware. Based on this proposal, \framework{} addresses the issues in current classically-inspired QCS methods and innovates in all important aspects of QCS with its topology-aware and data embedding-aware search space, noise-guided search algorithm, and two-stage circuit evaluation strategy. Comprehensive experimental
results show that \framework{}  significantly outperforms
leading QCS methods while achieving much more favorable resource efficiency and scalability. Due to its resource efficiency and improved search space, \framework{} provides a powerful tool for enabling new research in QML and noise-aware quantum compilation.

\section*{Acknowledgements}


S.A. thanks Sidharth Ramanan for fruitful discussions and comments on various drafts. Y.S. thanks Gushu Li for helpful reviews on the paper draft. We thank Prof. Devesh Tiwari for shepherding the paper and for providing detailed feedback. This work was funded in part by EPiQC, an NSF Expedition in Computing, under grant CCF-173044.
\ignore{
Here we elaborate on how we quantify the similarity between two representations for input data $x_i$ and $x_j$ using a circuit $C$, and how we use these similarities to compute each entry $R_C(i, j)$ of the matrix $R_C$. 

We define the \emph{representation} $\rho_C(x, \theta)$ of a data point $x$ when embedded in a circuit $C$ using parameters $\theta$ as the portion of the quantum state constructed by $C$ that corresponds to the $n_{\text{meas}}$ qubits being measured \emph{right before} the measurement takes place. A commonly employed method of computing similarity in the QML literature \cite{nature_communication_2021_power_of_data_qml, haug2021largescaleqml, schuld2021qmlmodelsarekernelmethods} is to use swap \cite{buhrman2001quantumfingerprinting, ngyuyen2021experimentalswaptest} or inversion test \cite{havlek2019suplearninginquantumenhancedfeaturespaces} circuits with pairs of quantum states, or, in this case, representations. However, these methods scale unfavorably, as the number of pairs to compute similarities for grows quadratically with the number of data points considered.

As a workaround, we use a randomized measurement protocol \cite{haug2021largescaleqml, huang2020classicalshadows, huang2022randomizedmeasurementtoolbox} to create approximations $\hat{\rho}_C(x, \theta)$ of representations $\rho_C(x, \theta)$ via a simple process outlined in~\cref{alg:candidategen}. We then use these classical approximations to compute pairwise similarities on a classical computer. This reduces the number of circuit executions required to linear in the number of data points, at the expense of increased classical computation. We compute similarities as

\begin{multline}
    \textrm{Sim}(\hat{\rho}_C(x_i, \theta), \hat{\rho}_C(x_j, \theta)) = \frac{1}{n_{\textrm{bases}}} \\ \sum_{k = 1}^{n_\textrm{bases}} 1 - \text{TVD}(\hat{\rho}_C(x_i, \theta)_k, \hat{\rho}_C(x_j, \theta)_k).    
\end{multline}

We define the average of $\textrm{Sim}(\hat{\rho}_C(x_i, \theta), \hat{\rho}_C(x_j, \theta))$ for two data points $x_i$ and $x_j$ over multiple $\theta$ as the \emph{induced similarity} $\textrm{IS}_C(x_i, x_j, \boldsymbol{\Theta} = \{\theta_1, \theta_2, \dots, \theta_{n_{\theta}}\})$ where $n_{\theta}$ is the number of different values for $\theta$ we average over:

\begin{equation}
    \textrm{IS}_C(x_i, x_j, \mathbf{\hat{\theta}}) = \frac{1}{n_{\theta}} \sum_{i = 1}^{n_{\theta}} \textrm{Sim}(\hat{\rho}_C(x_i, \theta), \hat{\rho}_C(x_j, \theta))
\end{equation}

If a circuit $C$ has high $\textrm{IS}_C(x_i, x_j, \boldsymbol{\Theta})$ for $x_i$, $x_j \text{s.t.} y_i = y_j$, and low $\textrm{IS}_C(x_i, x_j, \boldsymbol{\Theta})$ for $x_i$, $x_j \text{s.t.} y_i \neq y_j$, then the circuit on average constructs good representations for the data. Thus, it is likely to be easier for the parameter optimization process to find good $\theta$ for circuit $C$ than it would be for another circuit $C'$ which, on average, constructs bad representations for the data. Thus $C$ is likely to achieve better performance on the target dataset than $C'$.

\Metricone{}, denoted by $\text{Rep}(C)$ aims to capture the alignment between the labels of the input data and the induced similarities $\textrm{IS}_C(x_i, x_j, \boldsymbol{\Theta})$ in order to quantify the quality of the representations constructed by a circuit $C$. We do so via the process outlined \Sash{algo XX}, which arranges the induced similarities computed over a representative subset of the input data into a matrix $R_C$, and compares $R_C$ with a reference matrix $R_{\textrm{ref}}$ that captures the behaviour of an ideal circuit. The elements of $R_{\textrm{ref}}$ are set via

\begin{equation}
    R_{\textrm{ref}}(i, j) = \delta_{y_i = y_j} = \begin{cases}
  1 & \textrm{if}\:\:  y_i = y_j \\    
  0 & \textrm{otherwise.} 
  \end{cases}
\end{equation}

$\textrm{Rep}(C)$ is then computed as

\begin{equation}
    \textrm{Rep}(C) = \frac{d^2 - \| R_C - R_{\textrm{ref}}\|^{2}_{2}}{d^2}.
\end{equation}
}


\bibliographystyle{plain}
\bibliography{references}

\end{document}